# Exploring the reproducibility of functional connectivity alterations in Parkinson's Disease

Liviu Badea[1*], Mihaela Onu[2,3], Tao Wu[4,5], Adina Roceanu[6], Ovidiu Bajenaru[6,7]

1 Artificial Intelligence and Bioinformatics Group, National Institute for Research and Development in Informatics, Bucharest, Romania
2 Medical Imaging Department, Clinical Hospital Prof. Dr. Th. Burghele, 20 Panduri Street, Bucharest, Romania
3 University of Medicine and Pharmacy "Carol Davila", Biophysics Department, Bucharest, Romania
4 Department of Neurobiology, Key Laboratory on Neurodegenerative Disorders of Ministry of Education, Beijing Institute of Geriatrics, Xuanwu Hospital, Capital Medical University, Beijing, China
5 Beijing Key Laboratory on Parkinson's Disease, Parkinson Disease Centre of Beijing Institute for Brain Disorders, Beijing, China
6 University Emergency Hospital Bucharest, Neurology Department, Bucharest, Romania
7 University of Medicine and Pharmacy "Carol Davila", Department of Clinical Neurosciences, Bucharest, Romania
* Corresponding author

# Abstract

Since anatomic MRI is presently not able to directly discern neuronal loss in Parkinson's Disease (PD), studying the associated functional connectivity (FC) changes seems a promising approach toward developing *non-invasive* and *non-radioactive* neuroimaging markers for this disease. While several groups have reported such FC changes in PD, there are also significant discrepancies between studies. Investigating the reproducibility of PD-related FC changes on independent datasets is therefore of crucial importance.

We acquired resting-state fMRI scans for 43 subjects (27 patients and 16 normal controls, with 2 replicate scans per subject) and compared the observed FC changes with those obtained in two independent datasets, one made available by the PPMI consortium (91 patients, 18 controls) and a second one by the group of Tao Wu (20 patients, 20 controls).

Unfortunately, PD-related functional connectivity changes turned out to be non-reproducible across datasets. This could be due to disease heterogeneity, but also to technical differences. To distinguish between the two, we devised a method to directly check for disease heterogeneity using random splits of a single dataset. Since we still observe *non-reproducibility* in a large fraction of random splits of the same dataset, we conclude that functional heterogeneity may be a dominating factor behind the lack of reproducibility of FC alterations in different rs-fMRI studies of PD.

While *global* PD-related functional connectivity changes were non-reproducible across datasets, we identified a few *individual brain region pairs* with marginally consistent FC changes across all three datasets. However, training classifiers on each one of the three datasets to discriminate PD scans from controls produced only low accuracies on the remaining two test datasets. Moreover, classifiers trained and tested on random splits *of the same dataset* (which are technically homogeneous) also had low test accuracies, directly substantiating disease heterogeneity.

**Abbreviations**

PD Parkinson’s Disease

NC Normal Controls

rs-fMRI resting state functional Magnetic Resonance Imaging

FC Functional Connectivity

ROI Region of Interest

H&Y Hoehn and Yahr score

PPMI Parkinson’s Progression Markers Initiative

SVM Support Vector Machine classifier

GNB Gaussian Naïve Bayes classifier

# Introduction

Although Parkinson's disease (PD) is the second most common neurodegenerative disease after Alzheimer's disease, its diagnosis is still difficult, especially in the early premotor stages, as it is mainly based on clinical evidence. To date, there is still no unique standard diagnostic test for PD, despite the intense research efforts to develop accurate biomarkers based on blood tests or imaging scans. The best current objective tests for PD evaluate dopaminergic function in the basal ganglia by using various PET or SPECT radiotracers (e.g. DaTSCAN). But these tests make use of radioactive substances, are performed only in specialized imaging centers and can also be very expensive. Moreover, the loss of dopaminergic nigro-striatal neurons is a delayed pathological event in the evolution of the disease, corresponding to Braak stages III-IV.

On the other hand, conventional (CT or MRI) brain scans of PD patients usually appear normal or with minor non-specific changes, so that conventional imaging techniques are only useful for ruling out other diseases that can be secondary causes of parkinsonism.

Therefore, since anatomic MRI is presently not able to directly discern (dopaminergic) neuronal loss in PD [Tuite et al., 2013], studying the associated functional connectivity (FC) changes seems to be a promising approach toward developing *non-invasive* and *non-radioactive* neuroimaging markers for this disease.

While many groups have reported such FC changes in PD (see Supplementary Table 1 for a list of such studies), an in-depth analysis of existing literature revealed significant discrepancies between studies. Investigating the *reproducibility* of PD FC changes *on independent datasets* is therefore of crucial importance.

A comprehensive review and analysis of the literature related to resting-state fMRI studies of Parkinson's disease is out of the scope of the present paper [Gottlich et al., 2013; Long et al., 2012; Skidmore et al., 2013; Baudrexel et al., 2011; Wu et al., 2011; Kwak et al., 2010; Kwak et al., 2012; Helmich et al., 2010; Helmich et al., 2011; Luo et al., 2014; Yu et al., 2013; Hacker et al., 2012; Kurani et al., 2015; Baggio et al., 2015; Esposito et al., 2013; Szewczyk-Krolikowski et al., 2014; Tessitore et al., 2012; Sharman et al., 2013; Wu et al., 2012; Liu et al., 2013; Chen et al., 2015; Wen et al., 2013; Zhang et al., 2015] (Supplementary Table 1; see also the review by [Tahmasian et al., 2015]). We only mention some important inconsistencies of reported functional connectivity changes in PD. Due to the crucial importance of the striatum in PD, we first discuss some inconsistencies involving striatal seeds [Tahmasian et al., 2015]:

- Contrary to [Hacker et al. 2012], [Helmich et al. 2010] observed no significant difference in caudate functional connectivity in PD.
- On the other hand, contrary to the study [Helmich et al., 2010], [Luo, Song, et al. 2014] did not observe increased FC of the anterior putamen.
- In contrast to [Hacker et al., 2012], [Luo, Song, et al. 2014] did not find a FC decrease between the striatal seeds and the brainstem.

There are also inconsistencies involving non-striatal seeds. For example, [Wu et al., 2011] found disrupted FC between the pre-SMA and the left putamen, as opposed to [Helmich et al., 2010], who did not find a decreased FC between the putamen and pre-SMA in PD.

Since the motor symptoms are the most striking clinical manifestations in PD, many rs-fMRI studies of PD concentrate on the sensorimotor system, including the basal ganglia, while disregarding any other FC changes. On the other hand, other more unbiased studies tried to determine a more global picture of the FC changes in PD. Some even tried to develop classifiers for the disease based on rs-fMRI data [Long et al., 2012; Skidmore et al., 2013; Szewczyk-Krolikowski et al., 2014; Chen et al., 2015], but most studies were not validated on independent datasets.

There are also some gross discrepancies involving even the sign of the main FC changes in PD. For example, [Luo et al. 2014] found only decreased FC in early stage PD, whereas most studies also find FC increases.

The general picture one gets from the literature is complex and at times somewhat confusing due to the numerous inconsistencies. Of course, these inconsistencies could be due to the different disease stages analyzed, to the inherent functional heterogeneity of the disease, but also to technical differences, or to the differences in the complex data (pre)processing workflows. Therefore, it is crucial to use the same data processing workflow to check the reproducibility of PD-related FC changes on as many independent datasets as possible.

In this work, we report a comparison between three different datasets obtained by completely independent research groups (see Table 1). More precisely, we acquired resting-state scans for 43 Romanian subjects (27 patients and 16 normal controls, with 2 replicate scans per subject) and compared the observed functional connectivity changes with those obtained in two independent datasets, one made available by the PPMI consortium in the US (91 patients, 18 controls) and a second one by the group of Tao Wu in China (20 patients and 20 normal controls).

**Table 1. The 3 PD datasets compared in the present study (PD=Parkinson's Disease, NC=normal controls).**

| Dataset | PD subjects | NC subjects | PD scans | NC scans | age PD mean(SD) | age (NC) mean (SD) | p (age NC-PD) | Hoehn & Yahr mean(SD) | disease duration mean(SD) |
|---|---|---|---|---|---|---|---|---|---|
| NEUROCON | 27 (16 M) | 16 (5 M) | 54 | 31 | 68.7 (10.6) | 67.6 (11.9) | 0.76 | 1.92 (0.33) | 4.6 (6.5) |
| Tao Wu | 20 (11 M) | 20 (12 M) | 20 | 20 | 65.2 (4.4) | 64.8 (5.6) | 0.78 | 1.88 (0.63) | 5.4 (3.9) |
| PPMI | 91 (63 M) | 18 (14 M) | 134 | 19 | 61.3 (10.2) | 64.7 (9.7) | 0.17 | 1.72 (0.48) | 1.9 (1.0) |

We briefly describe our approach in Fig 1A to better guide the reader through the remainder of the paper. We started by comparing the PD-related global functional connectivity changes in the 3 datasets and found them to be non-reproducible. Of course, this could be due to disease heterogeneity, but also to technical differences. To better distinguish between these two possibilities, we devised a method to directly check for disease heterogeneity using random splits of a single dataset. On the other hand, we searched for *individual brain region pairs* with consistent connectivity changes across all three datasets. Finally, to more directly discriminate PD scans from controls, we trained multivariate machine learning classifiers on one dataset and tested them on the remaining two. Additionally, we trained and tested classifiers on technically homogeneous random splits of the same dataset, to more directly check for disease heterogeneity.

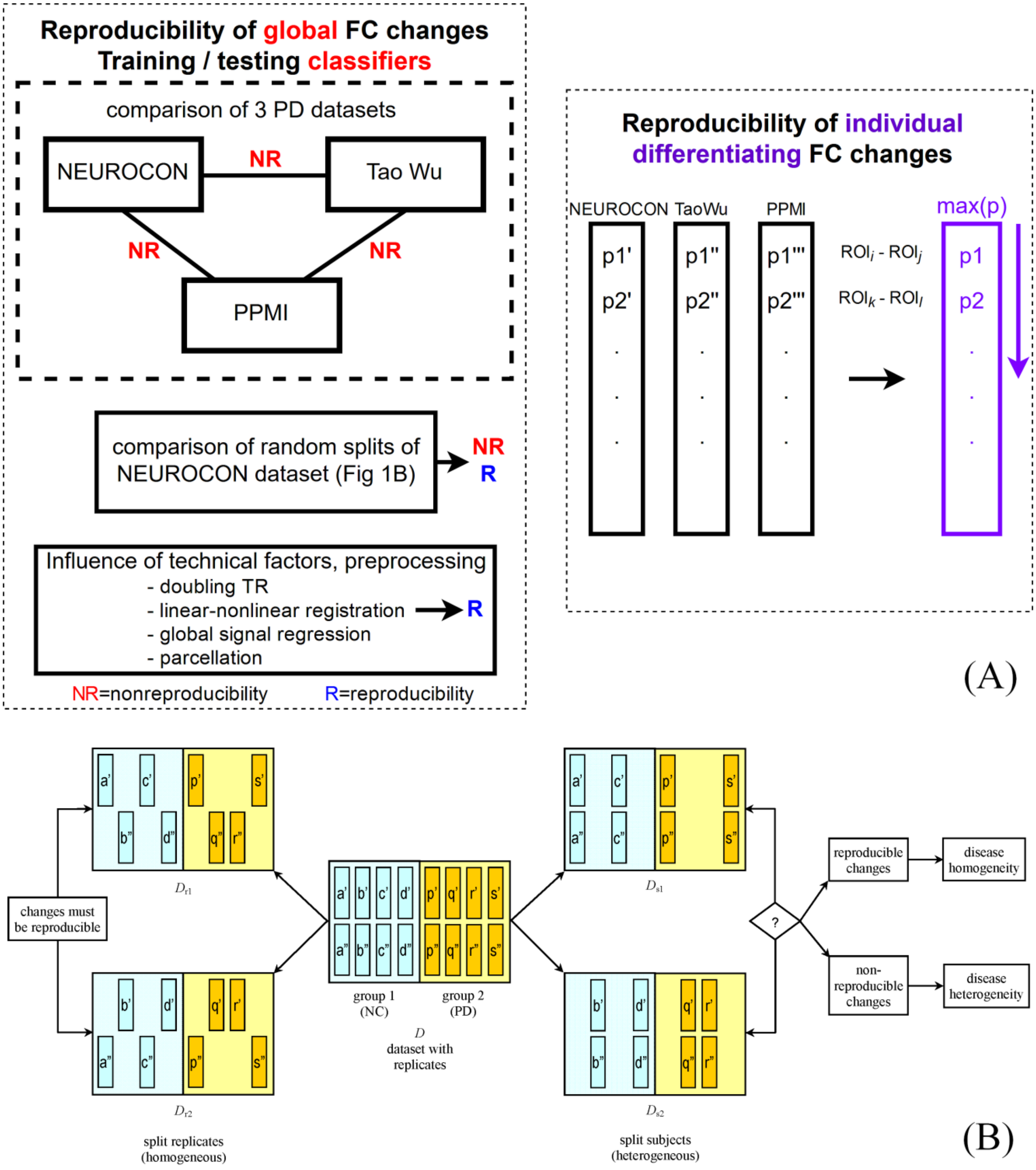

**Fig 1. Overview.** (A) Main steps of the analysis. (B) Using random splits of a dataset with replicate scans to check for disease (group) heterogeneity: (a, right) by placing different subjects (with all their replicate scans) in the two splits ("split subjects") and respectively (b, left) by splitting the replicates of the same subjects in the two splits ("split replicates"). The "split replicates" datasets must show reproducible changes anyway, while non-reproducible changes across the "split subjects" datasets are an indication of disease heterogeneity. (a,b,c,... correspond to subjects, while, for instance, a' and a" are replicate scans for subject a.)

This is the first study investigating the reproducibility of functional connectivity changes in Parkinson's disease on more than 2 datasets. Given the paucity of publicly available rs-fMRI PD datasets, we advocate the critical importance of data sharing for enabling the discovery of reproducible rs-fMRI biomarkers of PD.

# Materials and Methods

## *Datasets*

Three resting-state fMRI datasets of Parkinson's disease were compared in this study (see also Table 1 for the main patient characteristics and the numbers of scans in each study):

- the NEUROCON rs-fMRI study of 27 PD patients and 16 normal controls of the Neurology Department of the University Emergency Hospital Bucharest (Romania),
- a dataset of 20 PD patients and 20 normal controls provided by the group of Tao Wu (China),
- a publicly available dataset of 91 PD patients and 18 controls of the Parkinson's Progression Markers Initiative (PPMI) study in the US.

The datasets are somewhat similar, except for PPMI, which involved patients with a diagnosis of PD for two years or less and who are not taking PD medications, while most patients from the other two studies have been under treatment (most under levodopa). Also, PPMI patients were scanned in the 'eyes open' condition. Still, we argue that our findings were not affected by these differences. Since the datasets were compared in a pairwise manner, any putative discrepancies due to the shorter disease durations in the PPMI dataset would only show up in the NEUROCON-PPMI and Tao Wu-PPMI comparisons, but not in the NEUROCON-Tao Wu comparison. This was not observed in reality.

### NEUROCON

The NEUROCON study enrolled 27 patients with Parkinson's disease (mean age±SD 68.7±10.6 years) and 16 age-matched normal controls (67.6±11.9 years) with no history of neurological or psychiatric disease. The patients were clinically assessed at the Neurology Department of the University Emergency Hospital Bucharest (Romania) to be in the early or moderate stage of the disease according to the Queen Square Brain Bank (QSBB) clinical criteria and met the EFNS/MDS-ES (European Federation of Neurological Societies/Movement Disorder Society–European Section) recommendations for diagnosis of Parkinson's disease. The mean disease duration was 4.6 (±6.5) years for the entire patient cohort and respectively 2.75 (±2.15) years after excluding three patients with particularly long disease durations (over 10 years: 11, 16 and 32 years, respectively). Despite the longer disease durations, the above-mentioned 3 patients met the criteria for moderately advanced disease (H&Y stage 2) and thus were included in the study. The mean Hoehn and Yahr (H&Y) score [Hoehn and Yahr, 2001] was 1.93 (±0.33) and respectively 1.92 (±0.35) after excluding the 3 patients with long disease durations. All patients were in an early to moderate stage of disease (stages 1 to 2.5). The mean score on the motor subset of the Unified Parkinson's Disease Rating Scale (UPDRS) [Fahn, 1986] in the off medication condition was 28.3 (±9.3) for the entire patient cohort and 26.9 (±8.8) after excluding the 3 patients with long disease durations. The study has been approved by the University Emergency Hospital Bucharest ethics committee in accordance with the ethical standards of the 1964 Declaration of Helsinki and its later amendments. All patients gave their written informed consent to participate in the study.

**Scanning.** All subjects underwent two consecutive 8 min fMRI scans in a 1.5-Tesla Siemens Avanto MRI scanner, in an awake resting state with their eyes closed. Two consecutive replicate scans were acquired for each subject to enable the study of the reproducibility and respectively homogeneity of FC changes in PD. (A single (control) subject could only be scanned once.) The patients were scanned in the "off medication" state, at least 10 hours after the last intake of their medication. The scanning protocol involved an Echo Planar sequence with repetition time (TR) 3480 ms, echo time (TE) 50 ms, axial orientation, voxel size 3.8×3.8×5 mm (without slice gaps), flip angle 90° and number of averages=1. Each resting state session lasted 8.05 min, comprising 137 volumes. To enable better co-registration to the standard MNI template, high-resolution T1-weighted images were also obtained for all subjects using an MPRAGE sequence (IR method, TR=1940ms, TE=3.08ms, inversion time (IT)=1100ms, voxel size 0.97×0.97×1 mm, number of averages=1).

## Tao Wu

The dataset comprised 20 PD patients (11 males, mean age±SD 65.2±4.4 years) and 20 age-matched normal controls (12 males, 64.8±5.6 years). All patients were in the early to moderate stage of the disease (Hoehn and Yahr stages 1 to 2.5, except for a single patient with H&Y stage 3) and had normal Mini-Mental State Examination (MMSE) scores. Moreover, there was no statistically significant MMSE difference (p=0.43) between PD patients (28.8±1.1) and normal controls (29.1±1.3).

**Scanning.** Both resting state fMRI and anatomic T1 scans were acquired for the 40 subjects in a Siemens Magnetom Trio 3T equipment, in an awake resting state with their eyes closed. Each resting state session lasted about 8 min (239 volumes, TR=2s, TE=40 ms, flip angle=90°) with a voxel size of 4×4×5 mm (64×64 matrix, 28 slices, field of view=256mm×256mm). MPRAGE scans were also obtained (voxel size 1×1×1 mm) for registration to the MNI template.

## PPMI

Imaging data for 91 PD patients (63 males) and 18 normal controls (14 males) were downloaded from the Parkinson's Progression Markers Initiative (PPMI) study data portal (http://www.ppmi-info.org/access-data-specimens/download-data/, https://ida.loni.usc.edu/home/projectPage.jsp?project=PPMI) with the kind permission of the PPMI consortium. The study includes patients with a diagnosis of PD for two years or less and who are not taking PD medications. The patients and controls are age-matched (p=0.17, mean age±SD 61.3±10.2 years for the PD patients and respectively 64.7±9.7 years for the normal controls). The patients had a mean Hoehn & Yahr score of 1.72 (SD 0.48), with a mean disease duration (at the time of the scan) of 1.9 years (SD 1.0). All patients had H&Y scores 1 to 2, except for only two, who were classified as H&Y stage 3.

**Scanning.** The subjects were scanned in 8 different centers, but with a similar protocol on Siemens Tim Trio 3Tesla scanners. Each resting state session lasted about 8.4 min (210 volumes, TR=2.4s, TE=25ms, flip angle 80°) with a voxel size of 3.3×3.3×3.3 mm (68×66 matrix, 40 slices). Subjects were instructed to rest quietly, keeping their eyes open and not to fall asleep. MPRAGE scans were also obtained (voxel size 1×1×1 mm, TR=2.3s, TE=2.98ms, flip angle=9°) for registration to the MNI template.

Although our functional connectivity computations did not require any particular type of data normalization (as only inter-region correlations are computed, rather than amplitudes), we also considered a subset of scans acquired in a single center (center number 32, with the largest number of PD patient and normal control scans), referred to in the following by the suffix 'center32'.

## Preprocessing

All datasets were preprocessed in a uniform manner. The raw scanner data in DICOM format was converted to NIfTI using dcm2nii (https://www.nitrc.org/projects/dcm2nii/) and further preprocessed using FSL (FMRIB Software Library v5 http://fsl.fmrib.ox.ac.uk/fsl/) as follows: motion correction using MCFLIRT, brain extraction with BET, spatial smoothing (Gaussian kernel FWHM 5mm) and denoising using nonlinear filtering (SUSAN), temporal high-pass filtering (with a cutoff frequency of 0.01 Hz), registration to the standard Montreal Neurological Institute MNI152 template via the anatomical T1 image (more precisely, BBR registration of the BOLD image to the T1 image, followed by 12 DOF linear+nonlinear registration of the latter to the 2mm MNI template). Nonlinear registration was performed at a resampling resolution of 4mm.

Besides the above 'standard' preprocessing workflow, we also considered alternative workflows involving global signal regression (GS) and respectively a temporal bandpass filter (0.01-0.1Hz – an ideal low-pass 0.1Hz filter was used in addition to the default FSL 0.01Hz highpass filter).

Since subject motion in the scanner has been observed to have significant influence on the functional connectivities computed from rs-fMRI data, despite motion correction (e.g. [Power et al., 2015]), we also considered subsets of scans with low in-scanner motion (marked by the suffix '0', e.g. 'NC0' and 'PD0' – see also Supplementary Table 2).

# *PD-related functional connectivity changes*

*Functional connectivity* [Friston and Buchel, 2003] is a rather loosely defined term, which encompasses many different methods used to reveal temporal correlations of BOLD activity across the brain. The simplest method consists in computing the correlations between all pairs of regions of a given brain parcellation, but more sophisticated data decomposition methods, such as Independent Component Analysis (ICA) are also widely used. Such data decomposition methods do not assume a given brain parcellation, but instead construct spatial maps grouping voxels with highly correlated timecourses. (Still, instead of being given a parcellation, such methods need to be provided with a target number of components.)

In our study of the reproducibility of functional connectivity changes in PD, we used *brain parcellations constructed independently of the datasets under comparison*, rather than applying data decomposition methods, such as group-ICA, since the latter would be inherently biased towards the "training dataset". Group-ICA may obtain a better functional parcellation *for the training dataset*, but that parcellation would be less appropriate for any other independent dataset ("overfitting"), thereby introducing a bias in the analysis. To avoid these problems, we have chosen to use brain parcellations constructed independently of the datasets under comparison, including functional brain parcellations obtained by group-ICA on completely independent sets of subjects (such as the 'Stanford' functional parcellation [Shirer et al., 2011]).

Moreover, to compensate for potential biases of any specific parcellation, we extended our analyses to a number of 13 different parcellations employed in other rs-fMRI studies, two anatomical (AAL, Talairach) and 11 functional (see Table 2 for more details).

**Table 2. The brain parcellations used in the functional connectivity comparisons.**

| Parcellation | Reference | Number of regions | Comments |
|---|---|---|---|
| AAL | [Tzourio-Mazoyer et al., 2002] | 116 | anatomic atlas |
| Craddock 130 | [Craddock et al., 2012] | 130 | |
| Craddock 260 | | 260 | |
| Craddock 500 | | 500 | |
| Craddock 950 | | 950 | |
| Shen 100 | [Shen et al., 2013] | 93 | |
| Shen 200 | | 183 | |
| Shen 300 | | 278 | |
| OASIS | [Marcus et al., 2007] | 97 | |
| Power | [Power et al., 2011] | 264 | spherical regions with a 10mm radius |
| Gordon_surface | [Gordon et al., 2014] | 333 | |
| Talairach | [Talairach and Tournoux, 1988] | 695 | anatomic atlas |
| Stanford | [Shirer et al., 2011] | 90 | functional parcellation obtained by group-ICA |

For each parcellation, we computed average timecourses for each region of interest (ROI) and the *resting state functional connectivity* between each pair of ROIs ($ROI_1$,$ROI_2$) as the Fisher z-transform of the temporal correlation between the corresponding ROI timecourses:

$$FC(ROI_1,ROI_2) = z(corr(ROI_1,ROI_2)).$$

For each dataset, we determined significant PD-related FC changes by applying two-sample t-tests (with unequal sample sizes and unequal variances) to the functional connectivities of all ROI pairs. ROI-pairs with significant group differences (NC versus PD) represent regions whose functional connectivity was found to be significantly different in PD patients *in that particular dataset*. The main aim of this study is to determine whether these changes are reproducible across datasets, to enable the development of functional imaging biomarkers for PD.

# *Reproducibility of global functional connectivity changes in PD*

## Comparison of 3 different PD datasets

We first compared the *global PD-related functional connectivity changes* across the three independent datasets NEUROCON, Tao Wu and PPMI to check to what extent these changes are reproducible. More precisely, we performed pairwise comparisons for all dataset pairs as follows.

For each pair of datasets ($i$,$j$) and a fixed parcellation, we checked the extent to which the *PD-related FC changes* in one dataset are correlated to the changes in the second dataset.

*PD-related FC changes* were quantified using t-values $t(\mathrm{ROI}_k,\mathrm{ROI}_l)$ from group comparisons (unpaired two-sample t-tests between NC and PD) of the functional connectivities between pairs of regions of interest $FC(\mathrm{ROI}_k,\mathrm{ROI}_l)$.

Then the *reproducibility* $R_{ij}$ *across the two datasets i and j* was determined as the correlation between the corresponding t-values (viewed as a vector over all ROI pairs) for the two datasets:

$$R_{ij} = corr(T_i,T_j), \tag{1}$$

where

$$T_i=(t_i(\mathrm{ROI}_1,\mathrm{ROI}_2),\ t_i(\mathrm{ROI}_1,\mathrm{ROI}_3),\ t_i(\mathrm{ROI}_1,\mathrm{ROI}_4),\ \ldots)$$

with $t_i(\mathrm{ROI}_k,\mathrm{ROI}_l)$ the t-value corresponding to PD-related FC changes between $\mathrm{ROI}_k$ and $\mathrm{ROI}_l$ with respect to dataset $i$ (and similarly $T_j$ for dataset $j$).

For a more intuitive graphical depiction of reproducibility across two datasets, we also constructed the *scatter-plot of ROI-pair t-values* corresponding to group comparisons in the two datasets (see Fig 2 for an example of such a scatter-plot).

Comparing PD-related FC changes (t-values) in the two datasets amounts to plotting for each ROI-pair the t-value in dataset 1 against the t-value in dataset 2. We thereby obtain a scatter-plot with a point for each ROI pair. The comparison of the FC changes in the two datasets thus involves analyzing the distribution of points in the scatter-plot: ideally, perfect reproducibility would entail a diagonal distribution of points in the scatter-plot, corresponding to perfectly correlated t-values in the two datasets. Fig 4 depicts examples of good reproducibility, while Fig 2 shows cases of non-reproducibility across datasets.

The correlation of t-values for the two datasets $R_{ij} = corr(T_i,T_j)$, as introduced above in (1), can be viewed as an aggregate measure of the *reproducibility across the two datasets i* and *j*.

To obtain a more quantitative measure of the *statistical significance* of such a correlation $R_{ij}$ between datasets, we performed permutations of the group labels (NC and PD) *independently* for the two datasets and computed the *p-value of the* $R_{ij}$ *value* as the fraction of permutations $\sigma$ for which the dataset correlation w.r.t. the permuted data $R_{ij}^{(\sigma)}$ exceeds the real one ($R_{ij}$):

$$p_{ij} = |\{ \text{ permutation } \sigma \mid R_{ij}^{(\sigma)} \geq R_{ij} \}| / \mathrm{N}, \tag{2}$$

where $N$ is the total number of permutations. All our permutation tests involved $N$=1000 permutations.

Various factors have been mentioned in the literature to affect functional connectivity measures:

- subject motion in the scanner [Power et al., 2015],
- global signal regression (with or without) [Murphy et al., 2009; Hayasaka 2013],
- the choice of the parcellation.

To study the influence of these factors on our reproducibility results, we also considered subsets of scans with *low in-scanner motion* (marked by the suffix '0', e.g. 'NC0' and 'PD0'), repeated our analyses with global signal regression, bandpass filtering and performed the comparisons using all 13 brain parcellations previously mentioned. Since the PPMI data has been acquired in several different imaging centers, we also considered a potentially more homogeneous subset of scans acquired in a single center (center number 32, with the largest number of PD patient and normal control scans), referred in the following by the suffix ‘center32’.

## Comparison of random splits of the same PD dataset

As already mentioned in the Introduction, the observed lack of reproducibility of global FC changes across datasets could be due to disease heterogeneity, but also to technical differences. To distinguish between these two possibilities, we devised a method to directly check for disease heterogeneity using random splits of a single dataset with replicate scans. Technical differences can then be excluded since all the scans have been acquired under identical technical conditions. More precisely, since in the NEUROCON study we acquired two replicate scans for each subject, we constructed two *homogeneous dataset splits* simply by using (different) scans of the same subjects. Additionally, two *heterogeneous dataset splits* can be obtained by placing different subjects (with all their scans) in each split. In other words, instead of comparing two distinct datasets, we compared two random splits of the same dataset, either:

(a) by placing different subjects in the two splits, with all the replicate scans of a subject in the same split (“*split subjects*”, *heterogeneous split*), or

(b) by placing each replicate scan of the same subject in a different split, so that the two splits contain (different) scans of the same subjects (“*split replicates*”, *homogeneous split*).

Dataset splits (b) are homogeneous since they contain scans of the same subjects, while splits (a) are heterogeneous since they contain scans of different subjects. Therefore, *consistent reproducibility* across *all* random *heterogeneous splits* would indicate *disease homogeneity*, while *non-reproducibility* in a large fraction of random heterogeneous splits would imply *disease heterogeneity*. (In both cases, we expect to observe consistent reproducibility across the homogeneous splits, at least as long as the technical noise is not dominating the biological signal.) A diagram of our method is shown in Fig 1B.

As in the pairwise comparisons between different datasets, we used permutation tests and formula (2) to compute p-values of the reproducibility across split datasets, for both the heterogeneous (“split subjects”) and the homogeneous (“split replicates”) datasets. Due to the random nature of the splits, we repeated the analysis for $N_s > 1000$ different random splits of the original data. To assess the fraction of (non-)reproducible splits, we determined the empirical cumulative distribution function (CDF) of the reproducibility p-values for the $N_s$ random splits.

The analysis was also repeated for the data with global signal regression.

## Influence of technical factors, preprocessing and parcellation

We also studied the influence on reproducibility of certain key technical factors and preprocessing steps, such as:

- the repetition time (TR),
- linear vs. nonlinear registration,
- global signal regression,
- the specific brain parcellation used for evaluating functional connectivity.

The AAL parcellation (which is typical) was used whenever not specified otherwise.

### Doubling the TR

The repetition time might, in principle, influence the measured low-frequency rs-fMRI fluctuations and indirectly the functional connectivities (which are temporal correlations). To study the influence of the repetition time on reproducibility, we constructed a synthetic dataset with a double TR by leaving out every second time-point from the NEUROCON timeseries data (for each scan and each voxel). We then analyzed with our method the reproducibility of group changes in functional connectivity between the original NEUROCON dataset and the synthetic one with a double TR.

### Linear vs. nonlinear registration

To study the impact of registration on reproducibility, we preprocessed the NEUROCON data both with linear and nonlinear registration to the MNI 152 template and determined the reproducibility of group changes in functional connectivity between the two resulting datasets.

### Global signal regression

Since global signal regression (GSR) has been observed to be very effective at removing scanning artifacts [Hayasaka, 2013], including motion artifacts, we also studied the reproducibility of FC changes between the NEUROCON dataset processed with GSR and the same dataset processed without GSR.

### Influence of parcellation

To avoid potential biases of any specific parcellation, we repeated our pairwise comparisons between the 3 PD datasets using all 13 different parcellations mentioned above, including functional and anatomic parcellations, with a wide range of numbers of regions of interest (90 to 950).

## *Reproducibility of the individual differentiating FC changes*

The reproducibility analysis performed above involves *global* functional connectivity changes, i.e changes in the FC of *all* ROI pairs, not just the ones that differentiate PD from normal controls. Even with non-reproducible global FC changes, it might be in principle possible that only a very few brain region pairs might still reproducibly differentiate PD patients from controls. To this end, we also studied individual brain region pairs with FC changes that are significant w.r.t. *all* datasets.

More precisely, for each ROI-pair, we compute $max(p)$, the largest (least significant) of the three p-values obtained in the three datasets (separately for the FC increases and respectively decreases) and sort the ROI-pairs in increasing order of this $max(p)$. The most significant $min(max(p))$ of these $max(p)$ corresponds to the ROI-pair with the best overall significance with respect to the 3 datasets, as all other ROI-pairs have larger (less significant) p-values with respect to at least one dataset.

Finally, to assess the statistical significance of such a best ROI-pair, we use a permutation test (of the disease labels in each dataset) to check the fraction of random permutations with a more significant (smaller) $min(max(p))$ than the real data (we performed N=1000 random permutations).

$$p(\min(\max(p_+))) = |\{\text{ permutation } \sigma \mid min(max(p_+))^{(\sigma)} \leq min(max(p_+)) \}| / \text{N},$$

where $min(max(p_+))$ corresponds to FC increases in NC versus PD. A similar relation holds for the FC decreases $min(max(p_-))$.

The analysis was repeated for all 13 parcellations considered in this study (Table 2).

## *Learning classifiers for discriminating PD-related FC changes*

We used machine learning techniques to learn classifiers that discriminate PD from controls using functional connectivities between ROI pairs as features.

First, we trained classifiers on each one of the 3 datasets (NEUROCON, Tao Wu, PPMI) and tested them on the other two datasets. Both Linear Support Vector Machines (SVM) and Gaussian Naïve Bayes (GNB) classifiers were tested, with progressively increasing numbers of features: $N$=10,50,100,500,5000 (functional connectivities between ROI pairs). The $N$ features selected were the best discriminating ROI pair functional connectivities, based on unpaired t-tests between normal and PD scans. As the two classes (NC-Normal Controls and PD-Parkinson's Disease) are not balanced in all 3 datasets, we employ the *average accuracy Aacc* = (*acc*(NC)+acc(PD))/2 for assessing the performance of the classifiers (a random classifier is expected to have an average accuracy of 0.5).

Since the different datasets are not technically homogeneous, we also trained and tested classifiers on random splits of the same dataset, to check to what extent the low accuracies are due to technical differences, or to disease heterogeneity. More precisely, we performed 10,000 random splits in half of each dataset, trained a classifier on one half and tested it on the other.

# Results

ROI-pairs with significant group differences (NC versus PD) in functional connectivity were found in all three PD datasets: NEUROCON, Tao Wu and PPMI (Supplementary Tables 3 and 4). However, these changes seemed at first sight to be *distinct in each dataset*. Our main aim in this paper has been to systematically investigate the reproducibility of the PD-related FC changes across independent validation datasets.

## *PD-related FC changes are non-reproducible across 3 datasets*

The reproducibility of global PD-related functional connectivity changes was determined by pairwise comparisons between three independent datasets: NEUROCON, Tao Wu and respectively PPMI. Fig 2 shows the scatter-plots of ROI-pair t-values (corresponding to the group comparison NC-PD) for the three dataset pairs, indicating a lack of reproducibility of global FC changes in PD. (Perfect reproducibility would correspond to a diagonal distribution of points corresponding to ROI pairs with perfectly correlated t-values with respect to both datasets.) Moreover, discriminating ROI-pairs situated in the upper right and respectively lower left corners of one plot are not discriminating in the other plots.

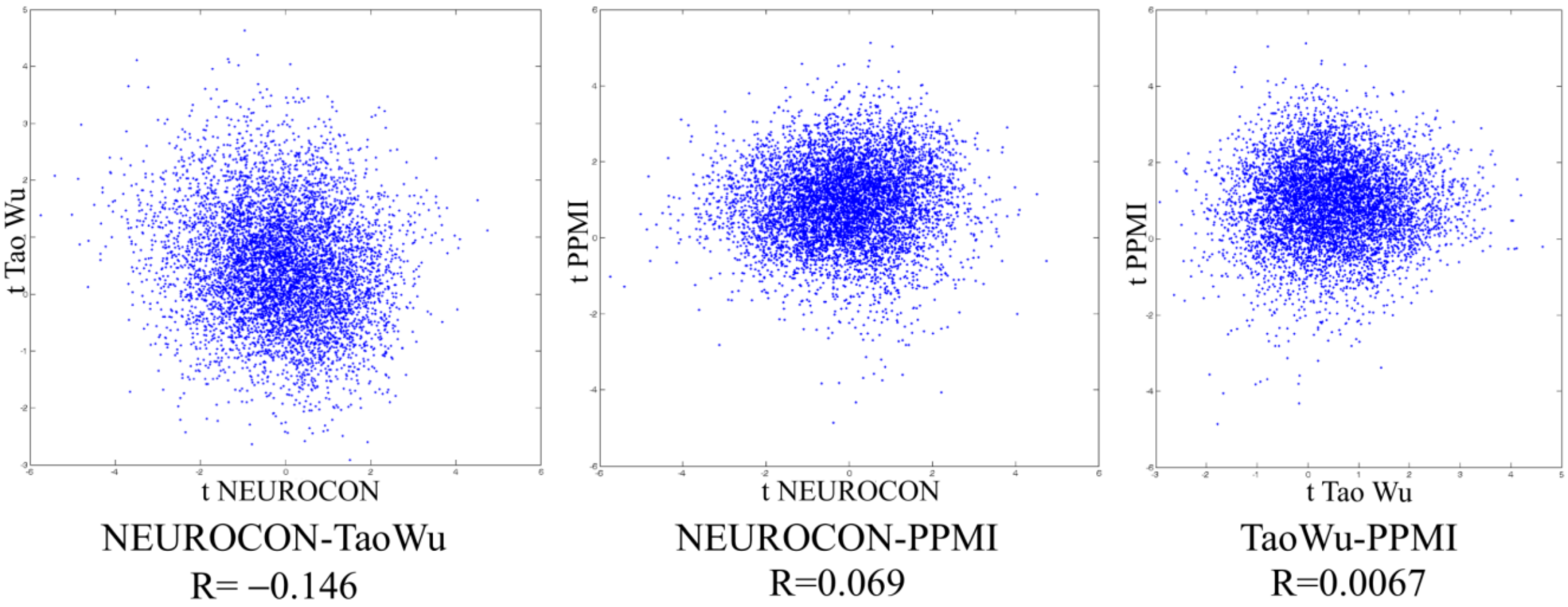

**Fig 2. Scatter-plots of ROI-pair t-values for the three dataset pairs indicate non-reproducibility of global PD-related FC changes.**

For a more quantitative measure of the reproducibility of FC changes between two datasets, we computed the Pearson correlation between t-values (viewed as vectors of over all ROI pairs) with respect to each dataset (*R* values shown in Fig 2.) We also estimated the statistical significance (p-values) of these reproducibility measures by permutation tests of the group labels independently for the two datasets – Table 3 shows the reproducibility measure and associated p-value for various pairwise comparisons between the three datasets, with standard highpass preprocessing (>0.01Hz), global signal regression (GS) and respectively bandpass filter (0.01-0.1Hz). Since in-scanner motion may influence FC measures, we present not only a comparison between the full patient and normal control cohort, but also that corresponding to a subset of scans with low in-scanner motion (denoted by the suffix '0'). Moreover, since PPMI data were acquired at many different centers, we also considered the restriction of the PPMI data to the scans from a single center (suffix 'center32'). The AAL parcellation was used in this case, but we also study the influence of the parcellation later on.

**Table 3. Reproducibility measure *R* and associated p-value *p* for various pairwise comparisons between datasets** with standard highpass preprocessing ('standard'), global signal regression (GS) and respectively bandpass filter (BP) ('0' indicates 'low in scanner motion', 'center32' – scans performed at a single PPMI center).

| Dataset 1 | Group contrast 1 | Dataset 2 | Group contrast 2 | standard | | GS | | BP | |
|---|---|---|---|---|---|---|---|---|---|
| | | | | *R* | *p* | *R* | *p* | *R* | *p* |
| NEUROCON | NC-PD | Tao Wu | NC-PD | -0.145955 | 0.949 | -0.0679698 | 0.804 | 0.000845579 | 0.476 |
| | NC0-PD0 | | NC0-PD0 | -0.0832455 | 0.847 | 0.035725 | 0.300 | 0.0132475 | 0.437 |
| NEUROCON | NC-PD | PPMI | NC-PD | 0.0692377 | 0.257 | 0.10039 | 0.138 | -0.163251 | 0.937 |
| | NC0-PD0 | | NC0-PD0 | 0.0401831 | 0.352 | 0.0308243 | 0.373 | -0.0824776 | 0.800 |
| | NC-PD | | NC_center32-PD_center32 | -0.0122797 | 0.526 | 0.0188561 | 0.430 | -0.142598 | 0.914 |
| | NC0-PD0 | | NC0_center32-PD0_center32 | -0.0873994 | 0.813 | -0.0894912 | 0.832 | -0.152545 | 0.945 |
| Tao Wu | NC-PD | PPMI | NC-PD | 0.00673472 | 0.466 | 0.0228885 | 0.403 | 0.0948706 | 0.153 |
| | NC0-PD0 | | NC0-PD0 | -0.0182896 | 0.560 | 0.044843 | 0.298 | 0.0545817 | 0.293 |
| | NC-PD | | NC_center32-PD_center32 | -0.0446155 | 0.687 | -0.032841 | 0.666 | 0.0243423 | 0.400 |
| | NC0-PD0 | | NC0_center32-PD0_center32 | -0.0712155 | 0.812 | -0.0176842 | 0.597 | -0.0436096 | 0.681 |

A clear lack of reproducibility of global PD-related FC changes is observed in all the three dataset pairs. This is the first study comparing three independent rs-fMRI datasets of PD. The fact that we compare 3 datasets is very important, as it lowers the probability that the lack of reproducibility is due to a dataset that may be "faulty" in some sense – in that case, with 3 datasets we might still observe reproducibility with respect to the remaining dataset pair (which we do not see in reality).

## *Inconsistent reproducibility of FC changes in heterogeneous dataset splits indicate disease heterogeneity*

The non-reproducibility across 3 datasets mentioned above seems to be due to disease heterogeneity, but it could also be due to technical differences. To exclude the latter possibility, we checked for disease heterogeneity using random splits of a single dataset with replicate scans (NEUROCON), all of which have been acquired under identical technical conditions.

(a) We first constructed random splits by placing different subjects in the two splits, with all the replicate scans of a subject in the same split ("*split subjects*", *heterogeneous splits*). As can be seen in Fig 3A (blue curve), a large fraction of these random splits display non-reproducible functional connectivity changes. More precisely, 88% of the random heterogeneous splits show non-reproducibility at the $p>0.01$ level and 42% at the $p>0.05$ level. Fig 3B shows the complementary cumulative distribution function (1-CDF) for the corresponding reproducibility measure *R* (blue curve), while Fig 3C presents a typical scatter-plot of ROI-pair t-values for a random heterogeneous split (one with *R* equal to the median).

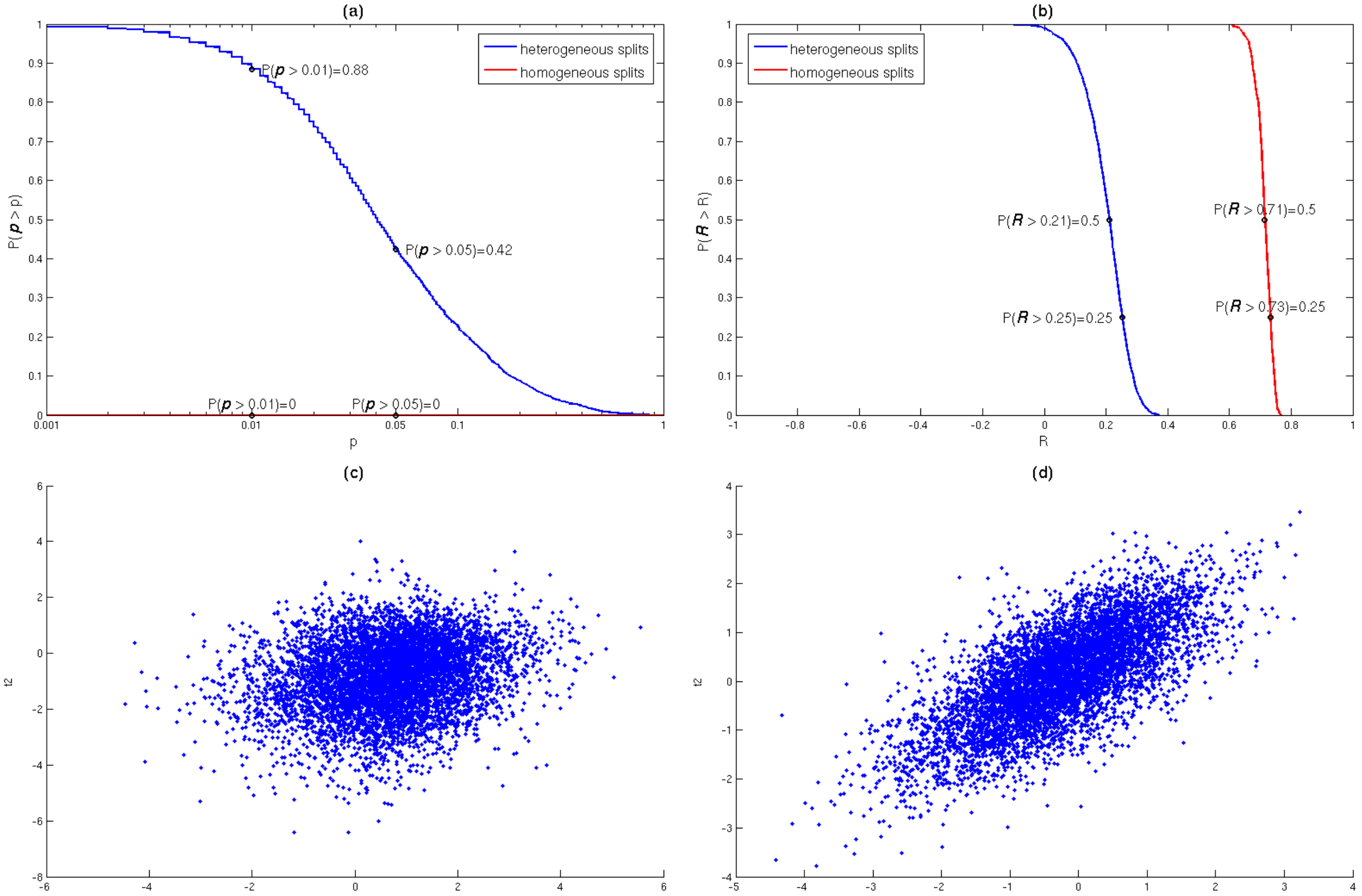

**Fig 3. *Inconsistent* reproducibility of PD-related FC changes in random *heterogeneous* dataset splits and *consistent* reproducibility in random *homogeneous* dataset splits.** (A) Complementary cumulative distribution function (CCDF=1-CDF) of the reproducibility p-values for $N_s$=2510 random *heterogeneous* splits and $N_s$=325 random *homogeneous* splits. (B) CCDF of the reproducibility measure *R*. (C) A scatter-plot of ROI-pair t-values for a random *heterogeneous* split. (D) A scatter-plot of ROI-pair t-values for a random *homogeneous* split.

(b) Next, we constructed random splits by placing each replicate scan of the same subject in a different split, so that the two splits contain (different) scans of the same subjects ("split replicates", homogeneous splits). In contrast to (a), all homogeneous splits showed reproducibility at the $p<10^{-3}$ level (Fig 3A, red curve). Fig 3B shows the complementary CDF for the reproducibility measure (red curve) – note the significantly higher reproducibility (*R*) values for the homogeneous splits (red curve, median *R*=0.71) as compared to the heterogeneous splits (blue curve, median *R*=0.21). Fig 3D displays a typical scatter-plot of ROI-pair t-values for a random homogeneous split.

The observed *non-reproducibility* in a large fraction of *heterogeneous dataset splits* indicates *disease heterogeneity*, in line with the comparison between the 3 independent PD datasets. As a control, we did indeed observe *consistent* reproducibility with respect to *all homogeneous* dataset splits, demonstrating that the technical noise could not have been the dominating factor behind the erratic non-reproducibility in heterogeneous splits.

The fact that the well-known *clinical* heterogeneity of Parkinson's disease is also accompanied by heterogeneity in resting state *functional connectivity* may not *retrospectively* be a big surprise to an experienced neurologist, although its exact extent could not have been estimated a priori, before analyzing the data. However, does this FC heterogeneity in PD also imply the lack of practical usefulness of rs-fMRI functional connectivity? Are there any other conditions that can be reliably differentiated using resting state functional connectivity? To answer these questions, we applied our approach to a different, potentially more homogeneous contrast, namely that between eyes open and eyes closed resting state conditions in healthy volunteers. Repeating our analysis of reproducibility of FC group changes on random splits of the Beijing eyes open-eyes closed dataset [Liu et al., 2013] (see Supporting Information) revealed *reproducibility* ($p<0.05$) not just in the *homogeneous* dataset splits, but also in the *heterogeneous* ones (Supplementary Fig 1 – only 6% of the heterogeneous and just 0.8% of the homogeneous random splits were non-reproducible at the $p>0.05$ level).

Summing up our findings, from the point of view of global FC changes, *Parkinson's disease* is *heterogeneous*, as opposed to the *eyes open-eyes closed* contrast, which is much more homogeneous (Table 4).

**Table 4. Summary of reproducibility of global functional connectivity changes in Parkinson's Disease and respectively the Eyes Open-Eyes Closed contrast.**

| **Reproducibility of global FC changes** | PD | EO-EC |
|---|---|---|
| different datasets | **no** | |
| split subjects (heterogeneous splits) | **inconsistent** (**no** in a large fraction of splits) | yes |
| split replicates (homogeneous splits) | yes | yes |

## *Influence of technical factors and preprocessing on reproducibility*

We found good reproducibility when changing various technical factors or processing options of the NEUROCON data, such as (see Fig 4):

- doubling the repetition time (TR),
- registration (linear versus nonlinear),
- global signal regression (with versus without).

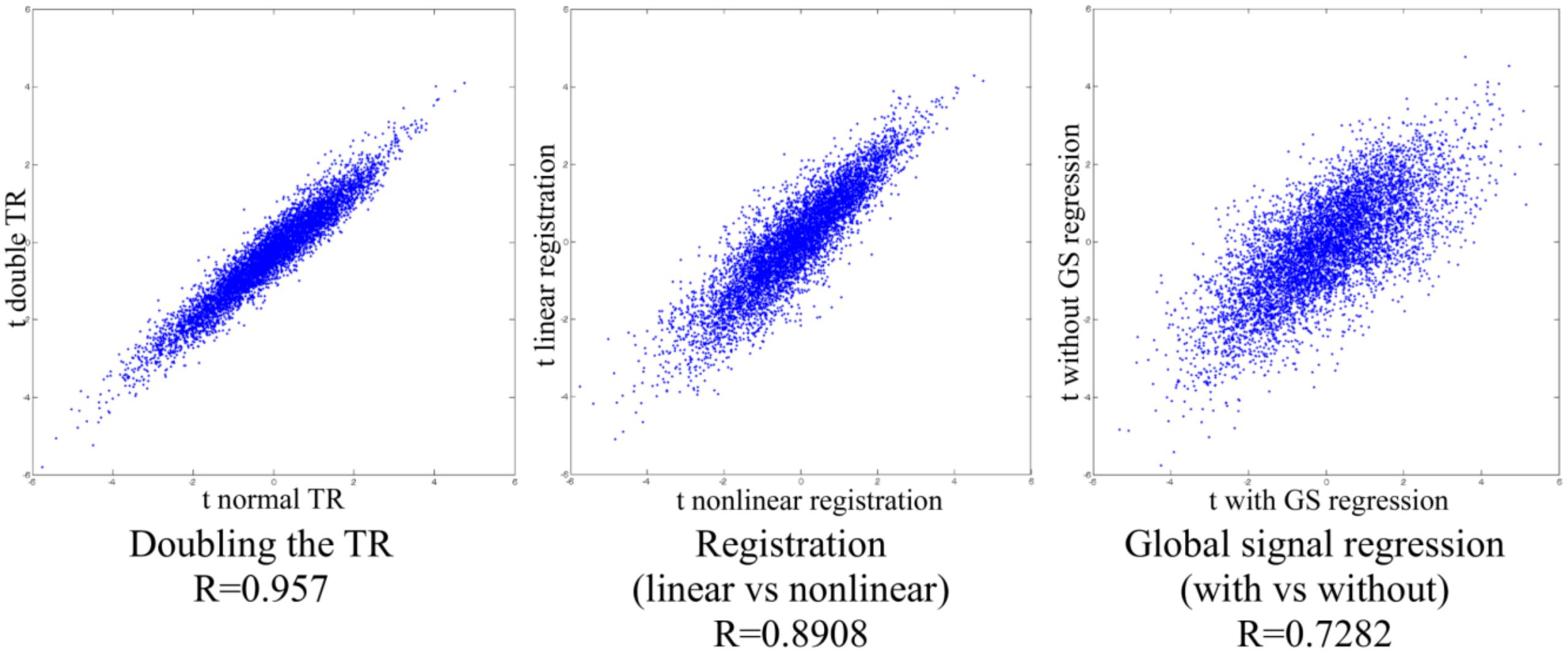

**Fig 4. Reproducibility when changing various technical factors or preprocessing options.**

This is in line with our conclusion that functional heterogeneity, rather than these technical factors, is the dominating factor behind the lack of reproducibility of FC changes in different rs-fMRI studies of Parkinson's disease.

We also tested the influence of various rs-fMRI denoising methods on the reproducibility of PD-related FC changes, such as ICA-FIX [Griffanti et al., 2014; Salimi-Khorshidi et al., 2014], or regression of the mean white matter and/or cerebrospinal fluid signal – none of these denoising methods changed the observed non-reproducibility (data not shown).

We also observed no improvement in reproducibility across random splits of the NEUROCON dataset after regressing out potential confounders, such as age, gender, or disease duration (data not shown).

## *Influence of parcellation on reproducibility*

We have argued that functional connectivity must be computed with respect to an unbiased parcellation (i.e. one that hasn't been constructed from any of the analyzed datasets). However, any given parcellation has also specific biases that may in principle affect the capacity to discriminate between PD and normal controls – an especially relevant factor is the average size and number of the ROIs. Testing the reproducibility of the PD-related global FC changes using 13 different parcellations, with varying numbers of ROIs (between 90 and 950, see Table 2), revealed a lack of reproducibility regardless of parcellation, or dataset pair (Table 5). (The NEUROCON-PPMI comparison was marginally significant (p=0.05) for the NC-PD contrast, but this significance didn't survive perturbations such as selecting just the 'center32' scans from PPMI (p=0.259), or restriction to the low motion scans NC0-PD0 (p=0.26), or NC0-PD0_center32 (p=0.555).)

**Table 5. Reproducibility measure and associated p-value for 13 parcellations and all three dataset pairs (NC-PD contrast).**

| Parcellation | NEUROCON-TaoWu | | NEUROCON-PPMI | | TaoWu-PPMI | |
|---|---|---|---|---|---|---|
| | R | p | R | p | R | p |
| Craddock130 | -0.011116 | 0.554 | 0.0927652 | 0.191 | -0.0613781 | 0.747 |
| Craddock260 | -0.00871389 | 0.560 | 0.0775174 | 0.205 | -0.0770479 | 0.864 |
| Craddock500 | 0.00488102 | 0.469 | 0.079955 | 0.172 | -0.0546894 | 0.839 |
| Craddock950 | 0.00952484 | 0.409 | 0.0694818 | 0.174 | -0.0347004 | 0.748 |
| Shen100 | -0.0490741 | 0.708 | 0.0958863 | 0.187 | -0.0567748 | 0.734 |
| Shen200 | -0.032232 | 0.641 | 0.0936098 | 0.172 | -0.0618746 | 0.799 |
| Shen300 | -0.0247647 | 0.632 | 0.0829943 | 0.159 | -0.0729234 | 0.871 |
| OASIS | 0.00698018 | 0.482 | 0.0205063 | 0.411 | -0.0155159 | 0.598 |
| Power264 | 0.0533032 | 0.166 | 0.0812484 | 0.141 | -0.0153593 | 0.600 |
| Gordon_surface | 0.0381087 | 0.262 | 0.025185 | 0.391 | -0.0435482 | 0.748 |
| Talairach | -0.0273596 | 0.686 | 0.0263891 | 0.357 | 0.0598013 | 0.129 |
| Stanford | -0.0600487 | 0.730 | 0.185859 | 0.050 | -0.081526 | 0.819 |
| AAL | -0.145955 | 0.949 | 0.0692377 | 0.257 | 0.00673472 | 0.466 |

## *Marginally significant individual differentiating FC changes in PD*

Despite non-reproducibility of PD-related *global* FC changes across different datasets, a small number of ROI-pairs that distinguish PD from controls may nevertheless, in principle, show reproducible changes across datasets. To check for this possibility, we concentrated on individual brain region pairs with FC changes that are significant w.r.t. *all* datasets, by sorting the ROI-pairs according to their least significance $max(p)$ with respect to *all* datasets.

For example, Table 6 shows the ROI-pairs with FC decreases in PD (i.e. positive t-values, corresponding to NC>PD) and max($p$) < 0.05 for the Power264 parcellation, without global signal regression. The best ROI-pair has max($p_+$)=0.0125, so $min(max(p_+))$=0.0125.

To check whether this $min(max(p_+))$ is statistically significant, we performed permutation tests as described. Table 7 lists these $min(max(p_\pm))$ values as well as their associated significance $p(min(max(p_\pm)))$ for all 13 parcellations. Only two out of the 13 parcellations yielded significant ROI-pairs at the p<0.05 significance level ('Power264' and 'Talairach'), while a third parcellation produced only marginally significant ROI-pairs ('Shen100', p=0.055) - see Table 8 and Fig 5. These (marginally) significant ROI-pairs involve visual-sensorimotor, respectively visual-parietal association areas. Whether these ROI-pair changes are more widely reproducible or not will have to await the release of more publicly-available PD rs-fMRI datasets.

**Table 6. Best ROI-pairs with FC decreases in PD (*t*>0, corresponding to NC>PD) and max(*p*) < 0.05 (for the Power264 parcellation, without global signal regression).**

| max(p) | ROI1-ROI2 | p NEUROCON (NC-PD) | p TaoWu (NC-PD) | p PPMI (NC-PD) | t NEUROCON (NC-PD) | t TaoWu (NC-PD) | t PPMI (NC-PD) |
|---|---|---|---|---|---|---|---|
| 0.012526 | sphere5(-21,-31,61)-sphere5(15,-77,31) | 0.012526 | 0.0018915 | 0.011638 | 2.5541 | 3.3398 | 2.7402 |
| 0.016354 | sphere5(-42,45,-2)-sphere5(43,-78,-12) | 0.0026347 | 0.016354 | 0.0079673 | 3.1172 | 2.5153 | 2.905 |
| 0.025311 | sphere5(-21,-31,61)-sphere5(-24,-91,19) | 0.0019939 | 0.025311 | 0.025232 | 3.2025 | 2.3298 | 2.3949 |
| 0.038357 | sphere5(-38,-33,17)-sphere5(20,-86,-2) | 0.038357 | 0.00041882 | 0.014314 | 2.1104 | 3.8665 | 2.653 |
| 0.038529 | sphere5(37,-81,1)-sphere5(38,-17,45) | 0.038529 | 0.02727 | 0.038269 | 2.1123 | 2.297 | 2.2003 |
| 0.038634 | sphere5(-21,-31,61)-sphere5(-26,-90,3) | 0.012064 | 0.038634 | 0.033842 | 2.5863 | 2.1425 | 2.2603 |
| 0.044406 | sphere5(-21,-31,61)-sphere5(29,-77,25) | 0.0085123 | 0.027058 | 0.044406 | 2.7193 | 2.3019 | 2.1243 |
| 0.046861 | sphere5(-21,-31,61)-sphere5(-40,-88,-6) | 0.0077414 | 0.046861 | 0.017283 | 2.7342 | 2.0557 | 2.55 |

**Table 7. Significance of *min*(*max*($p_\pm$)) values for all 13 parcellations (no global signal regression).**

| Parcellation | $min(max(p_+))$ | $min(max(p_-))$ | $p(min(max(p_+)))$ | $p(min(max(p_-)))$ |
|---|---|---|---|---|
| AAL | 0.0607935 | 0.112189 | 0.208 | 0.426 |
| Craddock130 | 0.0503968 | 0.242712 | 0.152 | 0.831 |
| Craddock260 | 0.0294203 | 0.148728 | 0.170 | 0.843 |
| Craddock500 | 0.0202304 | 0.0909226 | 0.206 | 0.859 |
| Craddock950 | 0.0172931 | 0.0542509 | 0.316 | 0.815 |
| Shen100 | 0.0371887 | 0.376232 | **0.055*** | 0.905 |
| Shen200 | 0.0425589 | 0.151199 | 0.195 | 0.747 |
| Shen300 | 0.0324906 | 0.115015 | 0.199 | 0.804 |
| OASIS | 0.0697703 | 0.3402 | 0.208 | 0.938 |
| Power264 | 0.012526 | 0.118716 | **0.033*** | 0.846 |
| Gordon_surface | 0.0275321 | 0.11095 | 0.185 | 0.846 |
| Talairach | 0.00701726 | 0.0339528 | **0.032*** | 0.522 |
| Stanford | 0.0599084 | 0.281706 | 0.145 | 0.820 |

**Table 8. Marginally significant FC changes w.r.t. all 3 datasets (decreased in PD).**

| Parcellation | $p(min(max(p_+)))$ | $min(max(p_+))$ | $ROI_1$ | $ROI_2$ |
|---|---|---|---|---|
| Talairach | 0.032 | 0.00701726 | (-24,-58,4)<br>left visual association area, lingual gyrus, BA18 | (–38, –32,16)<br>left superior temporal gyrus, BA41, planum temporale / parietal operculum |
| Power264 | 0.033 | 0.012526 | sphere5(-21,-31,61)<br>left postcentral / precentral gyrus | sphere5(15, –77,31)<br>right cuneus |
| Shen100 | 0.055 | 0.0371887 | L.BA19.3<br>left cuneus, precuneus | R.BA6.1<br>right SMA, middle cingulate |

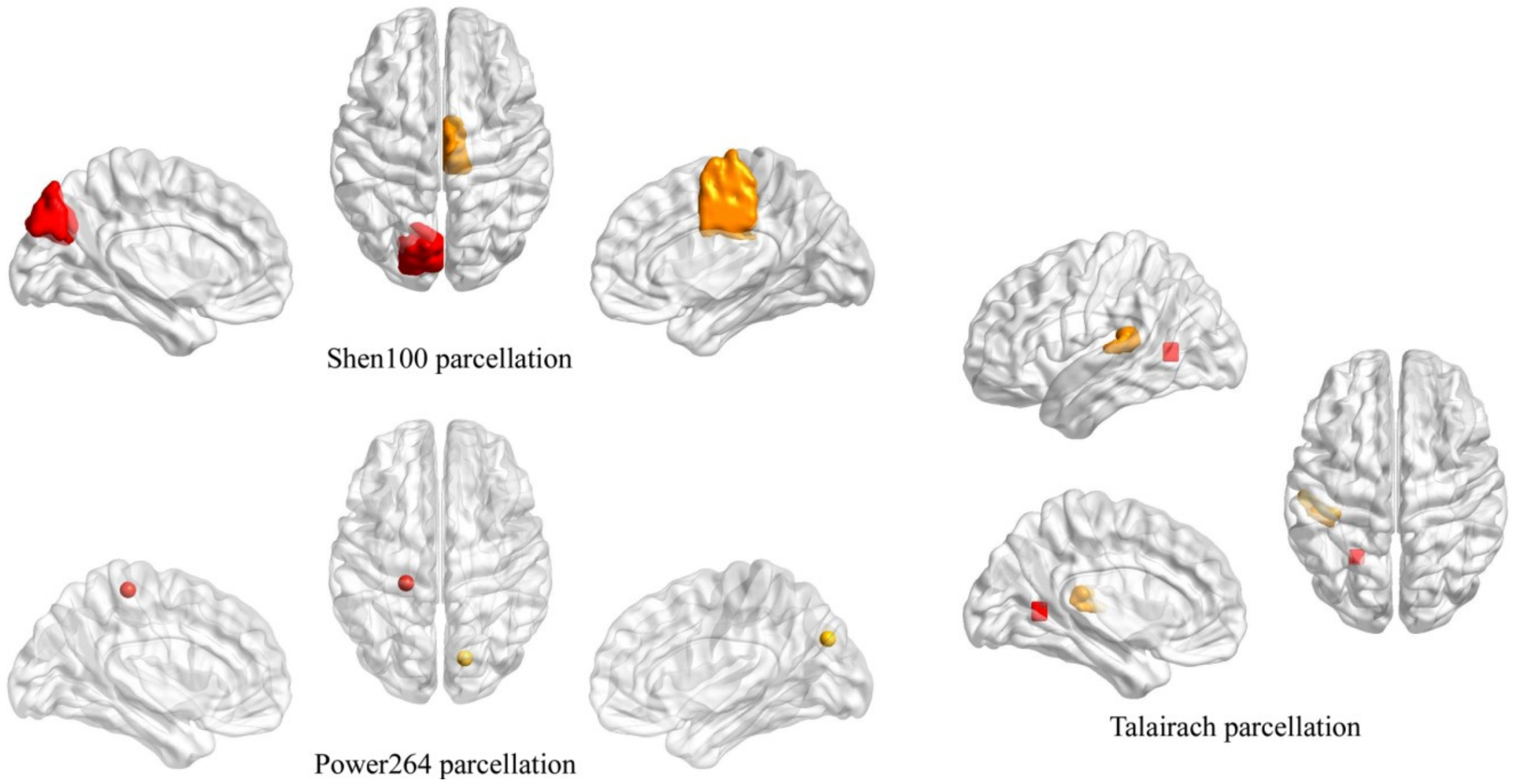

**Fig 5. Marginally significant FC changes w.r.t. all 3 datasets.** The ROIs were mapped onto the brain surface using BrainNet Viewer [Xia et al., 2013] (http://www.nitrc.org/projects/bnv/).

## *Testing classifiers for discriminating PD-related FC changes*

Training classifiers on functional connectivity data for each one of the 3 datasets (NEUROCON, Tao Wu, PPMI) and testing them on the other two datasets produced average accuracies on test data in the range 0.225 - 0.7 (mean 0.497, standard deviation 0.073), while a random classifier is expected to have an average accuracy of 0.5. Fig 6 shows the corresponding *average accuracies* $Aacc = (acc(NC)+acc(PD))/2$ for standard preprocessing (with the default FSL highpass filter at 0.01Hz), global signal regression and respectively bandpass filtering (0.01-0.1Hz) for both linear SVM and Gaussian Naïve Bayes (GNB) classifiers with $N$=5000 features (out of the total of 6670 ROI pairs of the AAL parcellation). See Supplementary Fig 2 and Supplementary Table 5 for the accuracies of classifiers with $N$=10,50,100,500,5000 features.

Since for each *training* dataset (for example NEUROCON), we have two different *test* datasets (PPMI and TaoWu in our example), we also computed an *aggregated* average accuracy for each dataset by taking the mean of the two average accuracies corresponding to the two remaining test datasets (Aacc(dataset-dataset1)+Aacc(dataset-dataset2))/2. The resulting aggregated average accuracies were low, in the range 0.336 - 0.591 (mean 0.497, standard deviation 0.0522, compared to 0.5 for a random classifier; see also Fig 7).

Since the three datasets are not technically homogeneous, we also trained and tested classifiers on random splits of the same dataset, to check to what extent the low accuracies are due to technical differences, or to disease heterogeneity. Fig 8 shows the average accuracies for 10,000 random splits in half of each dataset and various preprocessing options. Again, the means of the average accuracies over the 10,000 tests were low, in the range 0.51 – 0.66, reinforcing the evidence for disease heterogeneity.

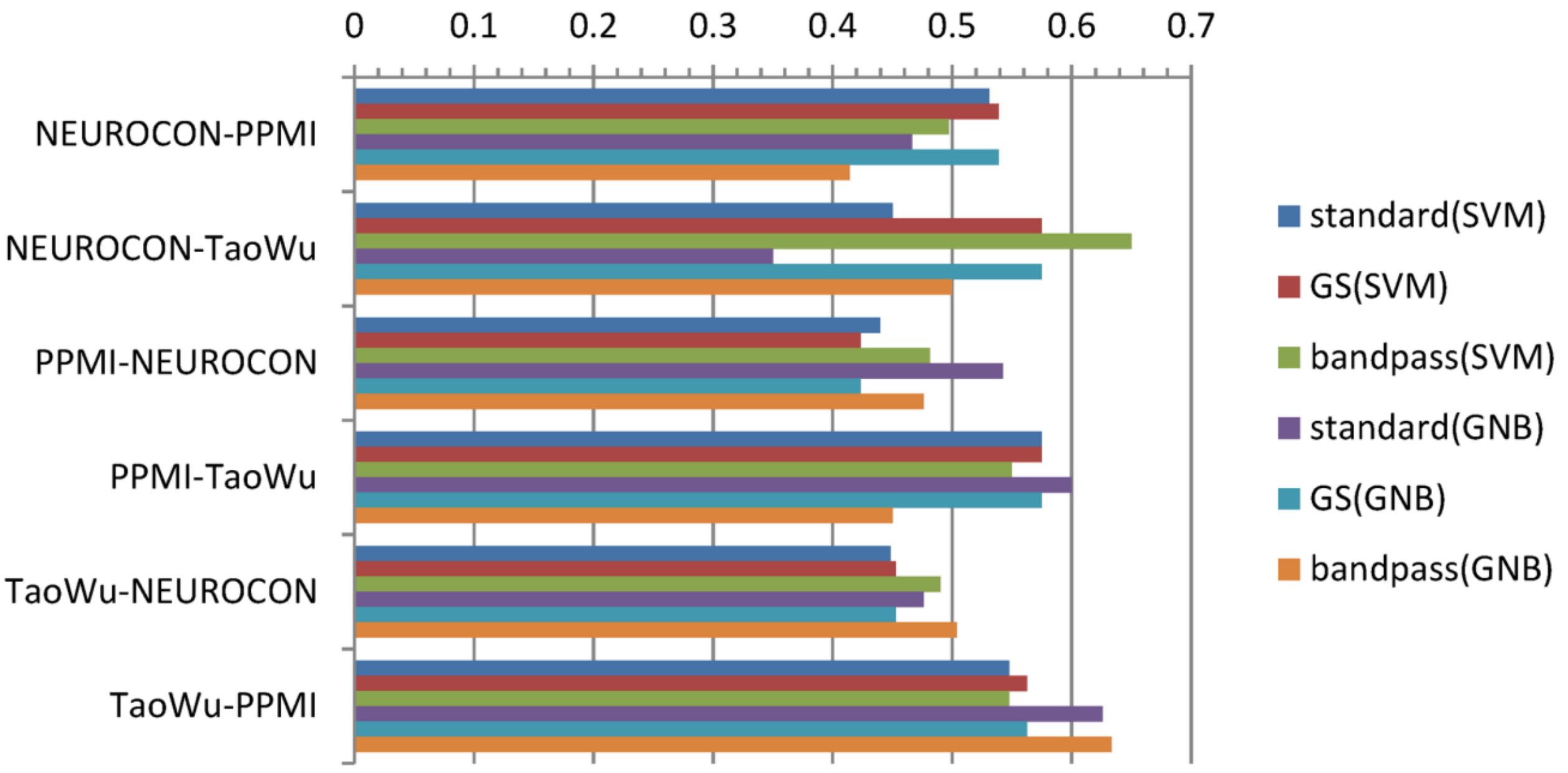

**Fig 6. Average accuracies for classifiers trained on dataset 1 and tested on dataset 2 for all dataset pairs** using standard preprocessing ('standard'), global signal regression (GS) and respectively bandpass filtering (0.01-0.1Hz). Here, SVM (linear Support Vector Machine) and GNB (Gaussian Naïve Bayes) classifiers used $N$=5000 features – see Supplementary Fig 2 for classifier accuracies for varying $N$. As an example, NEUROCON-PPMI denotes classifiers trained on NEUROCON and tested on PPMI data.

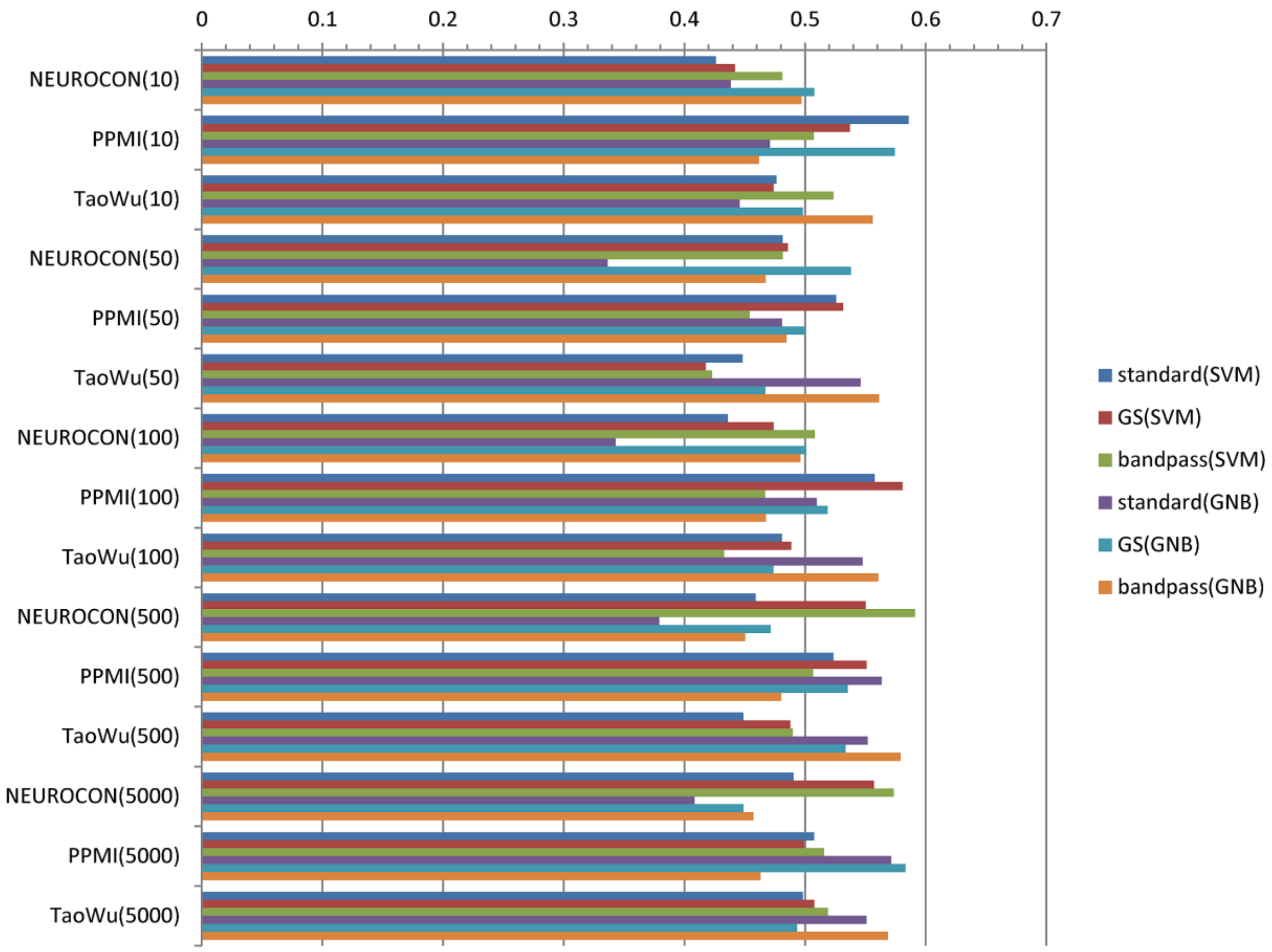

**Fig 7.** ***Aggregated*** **average accuracies for classifiers trained on each of the 3 datasets** using standard preprocessing ('standard'), global signal regression (GS) and respectively bandpass filtering (0.01-0.1Hz). Classifiers were trained with $N$=10,50,100,500,5000 features. As an example, NEUROCON(10) refers to the aggregated accuracy (Aacc(NEUROCON-PPMI) + Aacc(NEUROCON-TaoWu))/2 for classifiers trained on NEUROCON and tested on PPMI and respectively TaoWu data using $N$=10 features. SVM – linear SVM classifier, GNB – Gaussian Naïve Bayes classifier.

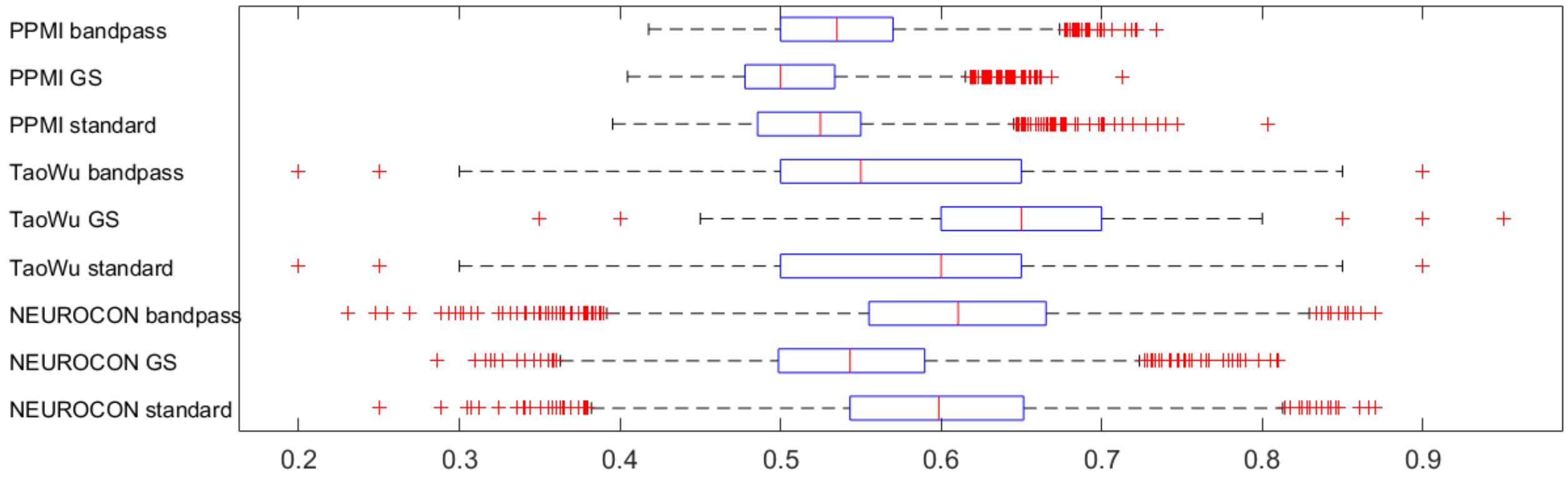

**Fig 8. Average accuracies for classifiers trained and tested on split data** ***from the same dataset*** using standard preprocessing ('standard'), global signal regression (GS) and respectively bandpass filtering (0.01-0.1Hz). An SVM classifier with $N$=5000 features was used.

# Discussion

The accelerated increase in the number of functional connectivity studies of Parkinson's Disease requires a consolidation of the knowledge in this field for enabling the development of clinically relevant rs-fMRI markers for this disease. Unfortunately however, there are many inconsistencies between published works and virtually no high confidence reproducibility studies.

This is the first study investigating the reproducibility of functional connectivity changes in Parkinson's disease on more than two datasets. The fact that we use a *uniform data processing workflow* for all datasets excludes a large number of technical factors as potential culprits for the observed differences between datasets. Also, the fact that our comparison involves *three* datasets is essential, as it lowers the probability that the observed lack of reproducibility is due to a problematic dataset – in such a case, with 3 datasets we might still observe reproducibility with respect to the remaining dataset pair, something which we do not see in reality.

To better clarify the issue, we devised a method to directly check for disease heterogeneity using random splits of a *single dataset* with replicate scans. Technical differences can then be excluded since all the scans have been acquired under identical technical conditions. The fact that we still observe non-reproducibility in a significant fraction of random subsamples of each individual dataset (these subsamples being *technically homogeneous* as they come from the same dataset), suggests that *functional heterogeneity* may be a dominating factor behind the lack of reproducibility of functional connectivity alterations in different resting state fMRI studies of Parkinson's disease.

This could be due to the heterogeneous multi-lesional topography and progression of the neurodegenerative process, possibly accompanied by variable compensatory functional circuit changes, as well as by changes due to dopaminergic medication [Tahmasian et al., 2015].

For a a more direct graphical depiction of the heterogeneity of the functional connectomes of the PD patients, we have applied *consensus NMF clustering* [Brunet et al., 2004] for a progressively increasing number of clusters ($k$=2,…,18, Supplementary Fig 3). Note that besides the consistent grouping of the replicate scan pairs for each patient, it is difficult to single out an optimal number of clusters $k$.

While *global* PD-related functional connectivity differences were non-reproducible across datasets, we identified a few *individual ROI pairs* with marginally consistent FC differences across all three datasets. However, finding out whether these differences are more widely reproducible or not will have to await the release of more public PD datasets.

Additionally, we applied more sophisticated multivariate machine learning techniques to learn classifiers that discriminate PD from controls using functional connectivities between ROI pairs as features. However, training classifiers on each one of the three datasets (NEUROCON, Tao Wu, PPMI) produced only low accuracies on the remaining two (test) datasets, in line with the preceding results. Furthermore, since the three datasets are not technically homogeneous, we also trained and tested classifiers on random splits *of the same dataset*, to more directly check to what extent the low accuracies are due to technical differences, or to disease heterogeneity. Again, we obtained low average accuracies (with means in the range 0.51 – 0.66), reinforcing the evidence for disease heterogeneity. Interestingly, these results are consistent with a recent study [Orban et al., 2017] on multisite generalizability of *schizophrenia* diagnosis based on functional brain connectivity, which reported multisite classification accuracies below 70%, in contrast to over 30 previously published, largely single-site schizophrenia studies, whose average reported classification accuracy exceeds 80%.

Therefore, given the paucity of publicly available rs-fMRI PD datasets, we advocate the critical importance of data sharing for enabling the discovery of reproducible and *clinically useful* functional imaging biomarkers of PD. In this regard, we view our study as an important first step towards more refined reproducibility studies that would be possible only with more publicly available datasets. In view of the many inconsistencies found in the published literature on PD-related functional connectivity changes, we strongly argue for a direct computational comparison of PD rs-fMRI datasets using a uniform data processing workflow, to avoid publication bias as well as processing workflow differences in the separate studies.

**Limitations.** The present study has concentrated on PD-related changes in *functional connectivity* (loosely viewed as *correlations* between different regions of interest), rather than changes in fluctuations of the *amplitude* of the rs-fMRI signal. In a complementary study, [Wu et al., 2015] observed PD-related changes in ALFF, but with rather limited reproducibility. An in-depth analysis of the reproducibility of PD-related differences in the *amplitude* of fluctuations is out of the scope of the present paper.

## Acknowledgments

We are very grateful to the PPMI consortium for granting us access to the Parkinson's Progression Markers Initiative (PPMI) imaging data. Data used in the preparation of this article were obtained from the PPMI database (www.ppmi-info.org/data). For up-to-date information on the study, visit www.ppmi-info.org. PPMI – a public-private partnership – is funded by the Michael J. Fox Foundation for Parkinson's Research and funding partners, including AbbVie, Avid Radiopharmaceuticals, Biogen, Bristol-Myers Squibb, Covance, GE Healthcare, Genentech, GlaxoSmithKline, Eli Lilly and Company, Lundbeck, Merck, Meso Scale Discovery, Pfizer, Piramal Imaging, Roche, Servier, UCB (Union Chimique Belge) and GOLUB Capital.

We also thank Madalina Tivarus and the reviewers for some very useful suggestions for improving the manuscript.

## Compliance with Ethical Standards

**Ethical approval:** The study was carried out in accordance with The Code of Ethics of the World Medical Association (Declaration of Helsinki) for experiments involving humans and was approved by the ethics committee of the University Emergency Hospital Bucharest.

**Informed consent:** Informed consent was obtained in written form from all individual participants included in the study.

# Supporting Information

## *Resting-state fMRI studies of Parkinson's disease*

**Supplementary Table 1.** Resting-state fMRI studies of Parkinson's disease

| Study | Main study characteristics | eyes open / closed | NC | PD | Patient characteristics |
|---|---|---|---|---|---|
| Gottlich 2013 | graphs, AAL, AAL subdivision in 343 cortical & subcortical ROIs | eyes closed | 20 | 37 | advanced disease, on medication (L-dopa, agonists) |
| Long 2012 | **classifier:** SVM (ALFF, ReHo, RFCS, GM, WM, CSF) accuracy 86.96%, sensitivity 78.95%, specificity 92.59%, precision 88% | eyes closed | 27 | 19 | early stage (H&Y 1-2), OFF (12 h) |
| Skidmore 2013 | **classifier** (ALFF): accuracy 88%, sensitivity 92%, specificity 87%, precision 87% | eyes closed | 15 | 14 | OFF(12-18 h) |
| Baudrexel 2011 | FC STN, M1 hand area | eyes closed | 44 | 31 | early stage PD patients (n=31) during the medication-off state with healthy controls (n=44); 16 tremor, 15 non-tremor |
| Wu 2011 | FC pre-SMA - M1<br>seeds: pre-SMA, M1, PCC | eyes closed | 18 | 18 | akinesia right (at most mild tremor), OFF (12h) H&Y 1.78+-0.5 |
| Kwak 2010 | 6 striatal seeds: 3 caudate (inferior ventral striatum, superior ventral striatum, dorsal caudate), 3 putamen (dorsal caudal putamen, dorsal rostral putamen, ventral rostral putamen) | eyes fixed on cross | 24 | 25 | mild to moderate stage (H&Y 1-2.5), ON and OFF(12–18 h) |
| Kwak 2012 | ALFF (fALLF) | | 24 | 24 | mild to moderate stage (H&Y 1-2.5) ON and OFF (12-18h) |
| Helmich 2010 | FC anterior, posterior putamen, caudate<br>posterior cingulate (control) | eyes closed | 36 | 41 | 13 PD without any tremor, 18 moderate to severe.<br>10 - never anti-PD medication (median H&Y 2.1, max 5)<br>perfectly age-matched (57 years) |
| Helmich 2011 | tremor | eyes closed | 36 | 41 | 19 tremor, 23 nontremor, right-handed, 12 no medication, OFF (12 h) |
| Luo 2014 | FC anterior, posterior putamen, anterior caudate, amygdala | eyes closed | 52 | 52 | 52 PD right handed, early stage drug-naïve, H&Y 1.85 (max 3), 31 right onset, 21 left onset |
| Yu 2013 | FC putamen, caudate, and SMA | | 20 | 19 | OFF<br>obvious at least a mild tremor |
| Hacker 2012 | FC striatum<br>6 ROIs: (i) caudate nucleus; (ii) anterior putamen; and (iii) posterior putamen | eyes open (fixate cross) | 19 | 13 | advanced PD **ON** |
| Kurani 2015 | FC (restricted to motor areas)<br>seeds: STN (most affected part), posterior cingulate (control seed) | eyes open (focused on the word "RELAX") | 19 | 39 | 20 de novo<br>19 moderate OFF |
| Baggio 2015 | cognitive impairement<br>ICA+dual regression (25 networks), group comparison of: DMN, dorsal attention network (DAN), bilateral frontoparietal networks (FPN),<br>43 seeds (10 DAN, 18 DMN, 15 FPN) | | 36 | 65 | ON state<br>65 nondemented: 34% mild cognitive impairement (MCI) |
| Esposito 2013 | ICA (fastICA BrainVoyager 40 components, restriction to sensorimotor network-best fit with previous template) | eyes closed | 18 | 20 | 20 drug naïve PD, 10 before & after (1H) levodopa, 10 before & after placebo |
| Szewczyk-Krolikowski 2014 | ICA+dual regression – basal ganglia (BG) network from 80 separate controls<br>- classifier based on average BG component values in voxels discriminating PD(OFF)-controls | eyes open | 19 | 19 | discovery cohort: 19 PD ON/OFF, 19 controls<br>validation cohort: 13 PD (5 drug-naïve), no controls (this disallows evaluation of specificity!)<br>80 elderly controls for BG template (MELODIC groupICA 50 components)<br>all subjects right-handed |
| Tessitore 2012 | fastICA, sogICA<br>40 components<br>select DMN, the frontoparietal (right and left FPN), sensorimotor network (SMN), visual, auditory | eyes closed | 15 | 29 | PD ON<br>16 FOG+ |
| Sharman 2013 | ROIs: caudate, putamen, globus pallidus, thalamus<br>sensorimotor (M1, postcentral gyrus)<br>associative (ventrolateral, dorsolateral prefrontal)<br>limbic (orbitofrontal, rectus gyrus, cingulate, insula, medial temporal ctx, perirhinal & entorhinal cortex, hippocampus, amygdala) | eyes closed | 45 | 36 | |
| Wu 2012 | ROIs: SNc bilaterally<br>effective connectivity Granger causality analysis (GCA)<br>with long TR=2000 & short TR=400 | eyes closed | 16 | 16 | 16 de novo<br>OFF/ON |
| Liu 2013 | FC dentate nucleus (cerebellum) -> cerebellar output | eyes closed | 18 | 18 | mild to moderate (1.34 H&Y average)<br>in OFF state<br>8 rigidity & bradykinesia-dominant (PD_AR)<br>10 tremor-dominant (PD_T) |
| Chen 2015 | SVM classifier LOOCV based on FC between 116 ROIs of AAL parcellation (150 features selected by Kendall tau correlation) - 93.62% accuracy, 90.47% sensitivity, 96.15 specificity | eyes closed | 26 | 21 | 21 PD (10 males, 11 females, 58.3 years), 26 HC (10 males, 16 females, 61.3 years) OFF state (12 hours) UPDRS 29.8 (sd 9.1), disease duration 3.2 years (sd 3.2) |
| Wen 2013 | depression<br>ALFF | eyes closed | 21 | 33 | OFF<br>17 depression |

### *Subject motion in scanner*

Since subject motion in the scanner has been observed to have significant influence on the functional connectivities computed from rs-fMRI data, despite motion correction (e.g. [Power et al., 2015]), we also considered subsets of scans with low in-scanner motion (marked by the suffix '0', e.g. 'NC0' and 'PD0' – see also Supplementary Table 2). The 3 datasets have different in-scanner motion characteristics.

**Supplementary Table 2.** The subsets of scans with low in-scanner motion

| Dataset | Low motion condition | NC scans | PD scans | NC0 scans | PD0 scans |
|---|---|---|---|---|---|
| NEUROCON | max abs motion ≤ 1/4 voxel = 0.958mm<br>max rel motion ≤ 1/6 voxel = 0.638mm | 31 | 54 | 27 | 39 |
| Tao Wu | max abs motion ≤ 1/4 voxel = 1mm<br>max rel motion ≤ 1/6 voxel = 0.67mm | 20 | 20 | 13 | 14 |
| PPMI | max abs motion ≤ 1/3 voxel = 1.098mm<br>max rel motion ≤ 1/3 voxel = 1.098mm<br>mean abs motion ≤ 1/6 voxel = 0.549mm<br>max rel motion ≤ 1/6 voxel = 0.549mm | 19 | 134 | 13 | 89 |

The following Tables show the p-values corresponding to potential group differences in motion in the 3 datasets analyzed:

**NEUROCON**

NC-PD (31-54) p-value (t-value)

| p(mean rel) | p(mean abs) | p(max rel) | p(max abs) |
|---|---|---|---|
| 0.00243021<br>(-3.14036) | 0.0545481<br>(-1.96039) | 0.0945346<br>(-1.6961) | 0.135713<br>(-1.50937) |

NC0-PD0 (27-39)

| p(mean rel) | p(mean abs) | p(max rel) | p(max abs) |
|---|---|---|---|
| 0.166962<br>(-1.39829) | 0.043621<br>(-2.06138) | 0.319776<br>(-1.00337) | 0.100148<br>(-1.67026) |

Note that in the NEUROCON cohort, patients move more than normal controls.

**Tao Wu**

NC-PD (20-20)

| p(mean rel) | p(mean abs) | p(max rel) | p(max abs) |
|---|---|---|---|
| 0.0270361 (2.33514) | 0.727106 (0.352521) | 0.0074933 (2.96994) | 0.543508 (0.613522) |

NC0-PD0 (13-14)

| p(mean rel) | p(mean abs) | p(max rel) | p(max abs) |
|---|---|---|---|
| 0.256895 (1.16397) | 0.497445 (0.688602) | 0.0859714 (1.80121) | 0.455533 (0.758043) |

In the Tao Wu cohort, patients move less than normal controls.

**PPMI**

NC-PD (19-134)

| p(mean rel) | p(mean abs) | p(max rel) | p(max abs) |
|---|---|---|---|
| 0.0619124 (1.9849) | 0.485967 (0.710965) | 0.107866 (1.69169) | 0.134632 (1.56558) |

NC0-PD0 (13-89)

| p(mean rel) | p(mean abs) | p(max rel) | p(max abs) |
|---|---|---|---|
| 0.0656489 (1.94269) | 0.314195 (-1.03452) | 0.224242 (1.26499) | 0.53616 (0.63367) |

NC_center32-PD_center32 (9-30)

| p(mean rel) | p(mean abs) | p(max rel) | p(max abs) |
|---|---|---|---|
| 0.242348 (1.23728) | 0.77672 (0.289486) | 0.344483 (1.00257) | 0.318376 (1.05993) |

NC0_center32-PD0_center32 (7-23)

| p(mean rel) | p(mean abs) | p(max rel) | p(max abs) |
|---|---|---|---|
| 0.173686 (1.40637) | 0.649951 (0.46271) | 0.252684 (1.21065) | 0.13755 (1.64485) |

## *ROI-pairs with significant group differences in the separate datasets*

Supplementary Table 3 below shows the numbers of significant ROI pairs for several significance thresholds for the unpaired t-test between patient and control functional connectivities (for the AAL parcellation and without correction for multiple comparisons, due to the limited sample sizes). A single scan for each subject was considered in this comparison.

**Supplementary Table 3.** Number of significant ROI-pairs for the AAL parcellation

| significance level | NEUROCON(16-27) | TaoWu(20-20) | PPMI(18-91) |
|---|---|---|---|
| 0.001 | 4 | 15 | 19 |
| 0.005 | 23 | 68 | 81 |
| 0.01 | 48 | 121 | 156 |
| 0.05 | 283 | 489 | 737 |

As expected, the power of the tests was greater for the datasets with more samples, but the effect sizes (t-values) were similar, as can be seen in Supplementary Table 4 below, which shows the top positive and respectively negative effect sizes observed.

**Supplementary Table 4.** Top effect sizes (t-values) and the corresponding p-values for the top positive and respectively negative FC alterations

| | NEUROCON(16-27) | TaoWu(20-20) | PPMI(18-91) |
|---|---|---|---|
| top $t_+$ | 3.05647 | 4.632511 | 4.70764 |
| top $p_+$ | 0.00429499 | 4.23E-05 | 7.99E-05 |
| top $t_-$ | -4.495779 | -2.91725 | -3.870344 |
| top $p_-$ | 6.09E-05 | 0.0062734 | 0.0003765 |

## *Comparison of random splits of an 'eyes open-eyes closed' dataset*

Besides the potentially heterogeneous contrast between PD and normal controls, we also tested our method on a different, potentially more homogeneous contrast, namely 'eyes open' versus 'eyes closed' resting state in healthy volunteers. We used the *Beijing eyes-open-eyes closed (EO-EC) dataset* [Liu et al., 2013] (http://fcon_1000.projects.nitrc.org/indi/IndiPro.html), which involved 48 college students aged 19–31 years, 24 female with no history of neurological and psychiatric disorders. Each participant underwent three 8 min resting state scanning sessions: an EC session followed by two sessions counter-balanced across subjects: one EO resting state and one EC resting state session.

The functional images were obtained on a Siemens Trio 3 Tesla scanner using an echo-planar imaging sequence with the following parameters: 33 axial slices, thickness/gap=3.5/0.7 mm, in-plane resolution=64×64, repetition time=2000 ms, echo time=30 ms, flip angle=90°, field of view (FOV)=200×200mm$^2$, 240 volumes per scan. In addition, a 3D T1-weighted MPRAGE image was acquired with the following parameters: 128 sagittal slices, slice thickness/gap=1.33/0 mm, in-plane resolution=256×192, TR=2530 ms, TE=3.39 ms, inversion time (TI)=1100 ms, flip angle=7°, FOV=256×256 mm$^2$. Note that the parameters used in this study are quite similar to the ones from the PD datasets, including the scanning time (~8 min), with the exception of the 1.5 Tesla field strength used in the NEUROCON study (all the other studies used 3 Tesla machines).

We repeated our analyses of reproducibility of group changes in functional connectivity on random splits of the Beijing EO-EC dataset on both "split subjects" (heterogeneous) and "split replicates" (homogeneous) datasets using the AAL parcellation. As in the case of PD, permutation tests were employed to compute p-values of the reproducibility across split datasets. Additionally, we repeated the analysis for the data with global signal regression.

## 'Eyes Open-Eyes Closed' FC changes are reproducible

The fact that the well-known *clinical* heterogeneity of Parkinson's disease is also accompanied by heterogeneity in resting state *functional connectivity* may not *retrospectively* be a big surprise to an experienced neurologist, although its exact extent could not have been estimated a priori, before analyzing the data.

However, does this FC heterogeneity in PD also imply the lack of practical usefulness of rs-fMRI functional connectivity? Are there any other conditions that can be reliably differentiated using resting state functional connectivity? To answer these questions, we applied our approach to a different, potentially more homogeneous contrast, namely that between eyes open and eyes closed resting state conditions in healthy volunteers. Repeating our analysis of reproducibility of FC group changes on random splits of the Beijing eyes open-eyes closed dataset [Liu et al., 2013] revealed *reproducibility* ($p<0.05$) not just in the *homogeneous* dataset splits, but also in the *heterogeneous* ones (Supplementary Fig 1 – only 6% of the heterogeneous and just 0.8% of the homogeneous random splits were non-reproducible at the $p>0.05$ level). This implies that the EO-EC contrast produces more homogeneous and reproducible global FC changes.

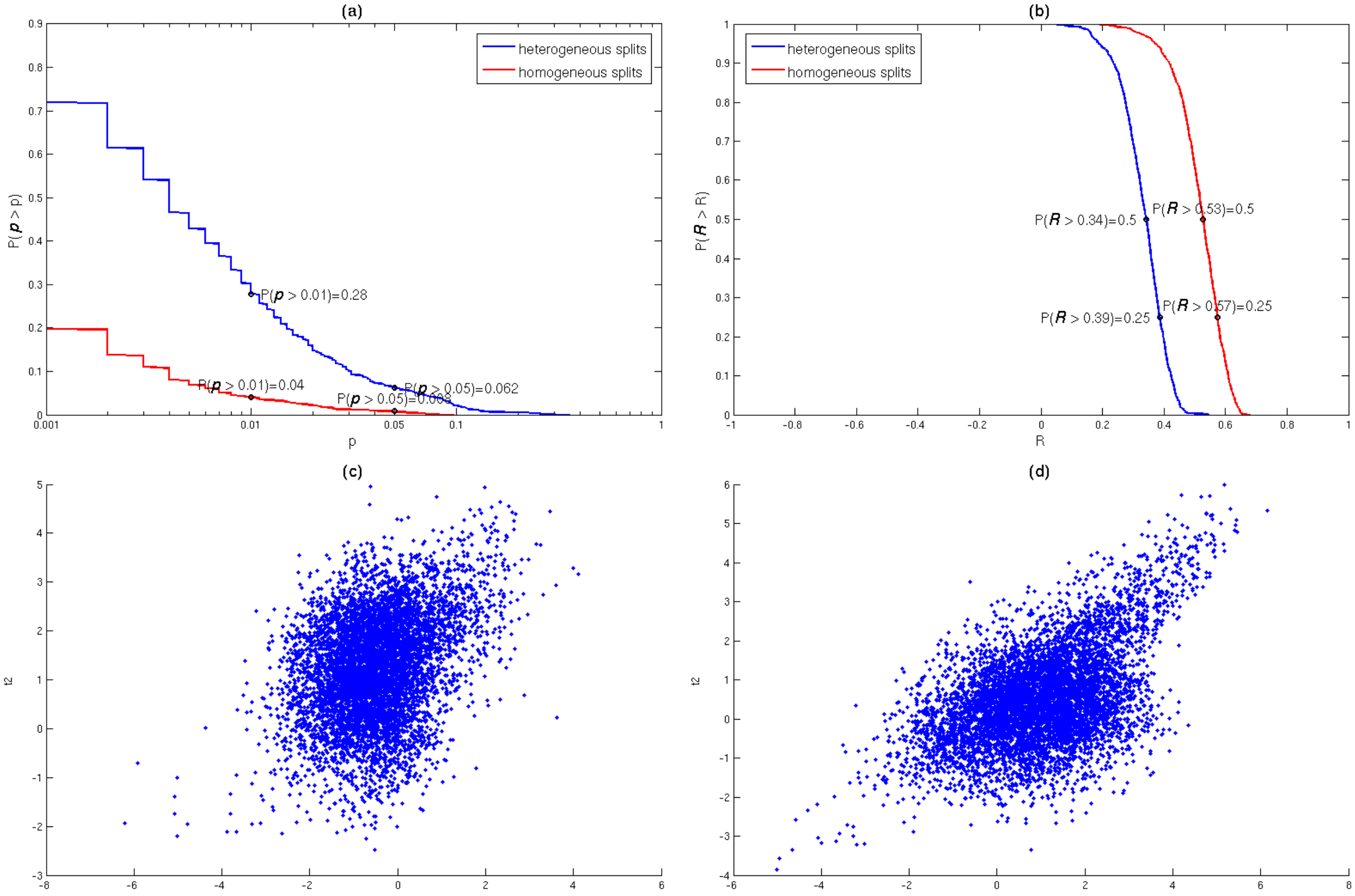

**Supplementary Fig 1. *Consistent* reproducibility of 'eyes open'-'eyes closed' (EO-EC) FC changes in random *heterogeneous* and *homogeneous* dataset splits.** (A) Complementary cumulative distribution function (CCDF=1-CDF) of the reproducibility p-values for $N_s$=1056 random *heterogeneous* splits and $N_s$=814 random *homogeneous* splits. (B) CCDF of the reproducibility measure *R*. (C) A scatter-plot of ROI-pair t-values for a random *heterogeneous* split. (D) A scatter-plot of ROI-pair t-values for a random *homogeneous* split.

## *Learning classifiers for discriminating PD-related FC changes*

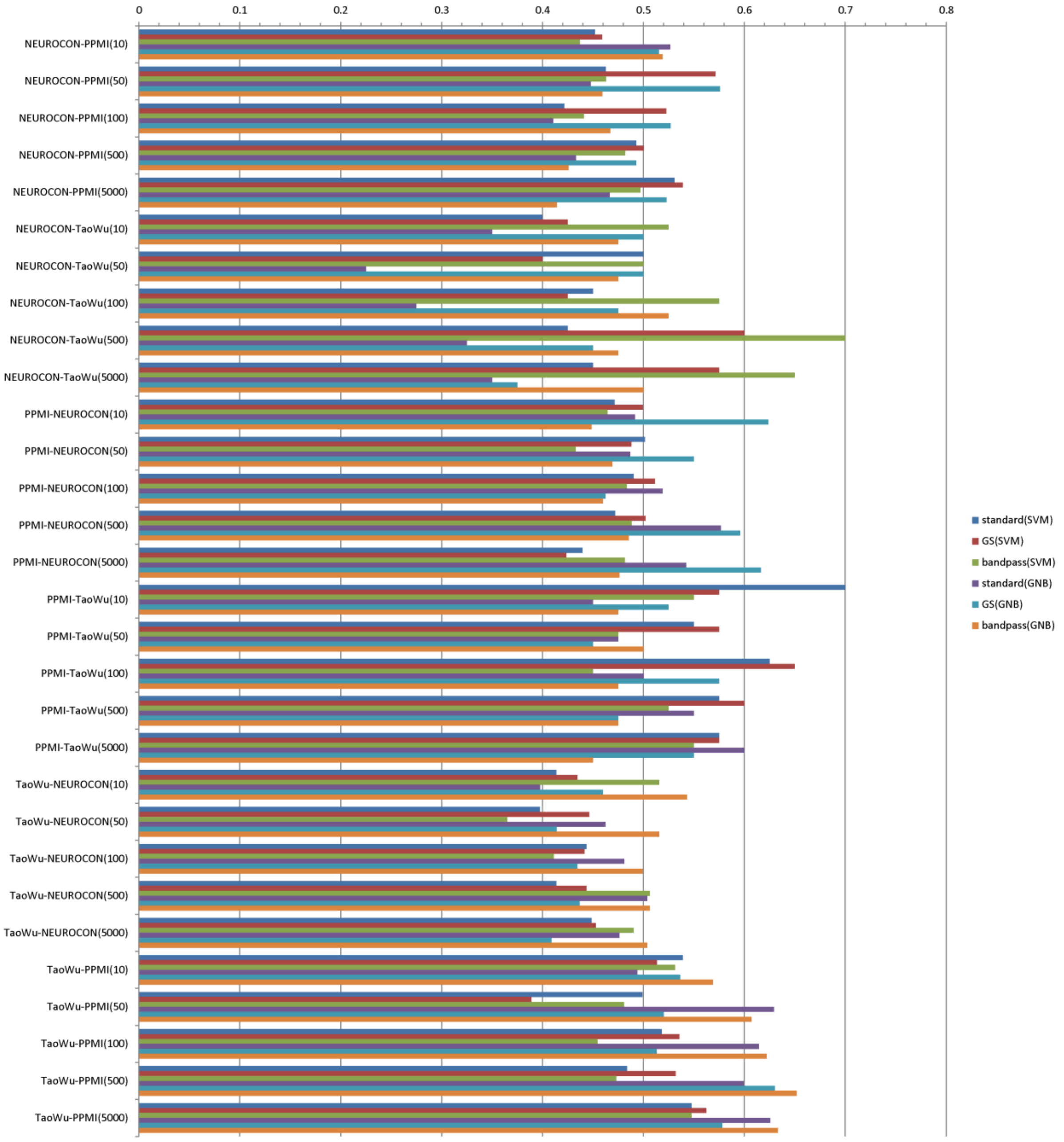

**Supplementary Fig 2. Average accuracies *Aacc* = (*acc*(NC)+acc(PD))/2 for classifiers trained on dataset 1 and tested on dataset 2 for all dataset pairs** using standard preprocessing ('standard'), global signal regression (GS) and respectively bandpass filtering (0.01-0.1Hz). Classifiers were trained with *N*=10,50,100,500,5000 features. For example, NEUROCON-PPMI(10) shows average accuracies of classifiers trained on NEUROCON and tested on PPMI data using *N*=10 features. SVM – linear SVM classifier, GNB – Gaussian Naïve Bayes classifier.

## Supplementary Table 5. Detailed performance metrics for the various classifiers

| TRAIN DATASET | TEST DATASET | classifier | N pairs | Pre-processing | TEST AvgaccPerClass | TEST accPerClass(NC) | TEST accPerClass(PD) |
|---|---|---|---|---|---|---|---|
| NEUROCON | PPMI | gnb_pooled | 10 | standard | 0.526709 | 0.1579 | 0.895522 |
| NEUROCON | PPMI | svm_linear | 10 | standard | 0.452082 | 0.1579 | 0.746269 |
| NEUROCON | PPMI | gnb_pooled | 50 | standard | 0.447761 | 0 | 0.895522 |
| NEUROCON | PPMI | svm_linear | 50 | standard | 0.462687 | 0 | 0.925373 |
| NEUROCON | PPMI | gnb_pooled | 100 | standard | 0.410644 | 0.05263 | 0.768657 |
| NEUROCON | PPMI | svm_linear | 100 | standard | 0.421642 | 0 | 0.843284 |
| NEUROCON | PPMI | gnb_pooled | 500 | standard | 0.433032 | 0.05263 | 0.813433 |
| NEUROCON | PPMI | svm_linear | 500 | standard | 0.49293 | 0.10526 | 0.880597 |
| NEUROCON | PPMI | gnb_pooled | 5000 | standard | 0.466614 | 0.05263 | 0.880597 |
| NEUROCON | PPMI | svm_linear | 5000 | standard | 0.530833 | 0.26316 | 0.798507 |
| NEUROCON | TaoWu | gnb_pooled | 10 | standard | 0.35 | 0.2 | 0.5 |
| NEUROCON | TaoWu | svm_linear | 10 | standard | 0.4 | 0.15 | 0.65 |
| NEUROCON | TaoWu | gnb_pooled | 50 | standard | 0.225 | 0.15 | 0.3 |
| NEUROCON | TaoWu | svm_linear | 50 | standard | 0.5 | 0.15 | 0.85 |
| NEUROCON | TaoWu | gnb_pooled | 100 | standard | 0.275 | 0.15 | 0.4 |
| NEUROCON | TaoWu | svm_linear | 100 | standard | 0.45 | 0.05 | 0.85 |
| NEUROCON | TaoWu | gnb_pooled | 500 | standard | 0.325 | 0.2 | 0.45 |
| NEUROCON | TaoWu | svm_linear | 500 | standard | 0.425 | 0.05 | 0.8 |
| NEUROCON | TaoWu | gnb_pooled | 5000 | standard | 0.35 | 0.25 | 0.45 |
| NEUROCON | TaoWu | svm_linear | 5000 | standard | 0.45 | 0.1 | 0.8 |
| PPMI | NEUROCON | gnb_pooled | 10 | standard | 0.491637 | 0.6129 | 0.37037 |
| PPMI | NEUROCON | svm_linear | 10 | standard | 0.471326 | 0.3871 | 0.555556 |
| PPMI | NEUROCON | gnb_pooled | 50 | standard | 0.486858 | 0.67742 | 0.296296 |
| PPMI | NEUROCON | svm_linear | 50 | standard | 0.501493 | 0.35484 | 0.648148 |
| PPMI | NEUROCON | gnb_pooled | 100 | standard | 0.519116 | 0.74194 | 0.296296 |
| PPMI | NEUROCON | svm_linear | 100 | standard | 0.490143 | 0.25807 | 0.722222 |
| PPMI | NEUROCON | gnb_pooled | 500 | standard | 0.576762 | 0.83871 | 0.314815 |
| PPMI | NEUROCON | svm_linear | 500 | standard | 0.471924 | 0.12903 | 0.814815 |
| PPMI | NEUROCON | gnb_pooled | 5000 | standard | 0.542413 | 0.67742 | 0.407407 |
| PPMI | NEUROCON | svm_linear | 5000 | standard | 0.439665 | 0.06452 | 0.814815 |
| PPMI | TaoWu | gnb_pooled | 10 | standard | 0.45 | 0.85 | 0.05 |
| PPMI | TaoWu | svm_linear | 10 | standard | 0.7 | 0.6 | 0.8 |
| PPMI | TaoWu | gnb_pooled | 50 | standard | 0.475 | 0.8 | 0.15 |
| PPMI | TaoWu | svm_linear | 50 | standard | 0.55 | 0.3 | 0.8 |
| PPMI | TaoWu | gnb_pooled | 100 | standard | 0.5 | 0.85 | 0.15 |
| PPMI | TaoWu | svm_linear | 100 | standard | 0.625 | 0.35 | 0.9 |
| PPMI | TaoWu | gnb_pooled | 500 | standard | 0.55 | 0.9 | 0.2 |
| PPMI | TaoWu | svm_linear | 500 | standard | 0.575 | 0.15 | 1 |
| PPMI | TaoWu | gnb_pooled | 5000 | standard | 0.6 | 0.85 | 0.35 |
| PPMI | TaoWu | svm_linear | 5000 | standard | 0.575 | 0.15 | 1 |
| TaoWu | NEUROCON | gnb_pooled | 10 | standard | 0.397551 | 0.25807 | 0.537037 |
| TaoWu | NEUROCON | svm_linear | 10 | standard | 0.41368 | 0.29032 | 0.537037 |
| TaoWu | NEUROCON | gnb_pooled | 50 | standard | 0.462366 | 0.25807 | 0.666667 |
| TaoWu | NEUROCON | svm_linear | 50 | standard | 0.397252 | 0.3871 | 0.407407 |
| TaoWu | NEUROCON | gnb_pooled | 100 | standard | 0.480884 | 0.25807 | 0.703704 |
| TaoWu | NEUROCON | svm_linear | 100 | standard | 0.443548 | 0.3871 | 0.5 |
| TaoWu | NEUROCON | gnb_pooled | 500 | standard | 0.503883 | 0.32258 | 0.685185 |
| TaoWu | NEUROCON | svm_linear | 500 | standard | 0.41368 | 0.29032 | 0.537037 |
| TaoWu | NEUROCON | gnb_pooled | 5000 | standard | 0.476105 | 0.32258 | 0.62963 |
| TaoWu | NEUROCON | svm_linear | 5000 | standard | 0.448626 | 0.19355 | 0.703704 |
| TaoWu | PPMI | gnb_pooled | 10 | standard | 0.493912 | 0.36842 | 0.619403 |
| TaoWu | PPMI | svm_linear | 10 | standard | 0.539081 | 0.47368 | 0.604478 |
| TaoWu | PPMI | gnb_pooled | 50 | standard | 0.629419 | 0.68421 | 0.574627 |
| TaoWu | PPMI | svm_linear | 50 | standard | 0.499018 | 0.73684 | 0.261194 |
| TaoWu | PPMI | gnb_pooled | 100 | standard | 0.614493 | 0.68421 | 0.544776 |
| TaoWu | PPMI | svm_linear | 100 | standard | 0.518068 | 0.84211 | 0.19403 |
| TaoWu | PPMI | gnb_pooled | 500 | standard | 0.599568 | 0.68421 | 0.514925 |
| TaoWu | PPMI | svm_linear | 500 | standard | 0.483896 | 0.68421 | 0.283582 |
| TaoWu | PPMI | gnb_pooled | 5000 | standard | 0.625687 | 0.68421 | 0.567164 |
| TaoWu | PPMI | svm_linear | 5000 | standard | 0.547722 | 0.78947 | 0.30597 |
| NEUROCON | PPMI | gnb_pooled | 10 | GS | 0.515318 | 0.10526 | 0.925373 |
| NEUROCON | PPMI | svm_linear | 10 | GS | 0.458955 | 0 | 0.91791 |
| NEUROCON | PPMI | gnb_pooled | 50 | GS | 0.575805 | 0.31579 | 0.835821 |
| NEUROCON | PPMI | svm_linear | 50 | GS | 0.571485 | 0.1579 | 0.985075 |
| NEUROCON | PPMI | gnb_pooled | 100 | GS | 0.526905 | 0.21053 | 0.843284 |
| NEUROCON | PPMI | svm_linear | 100 | GS | 0.522584 | 0.05263 | 0.992537 |
| NEUROCON | PPMI | gnb_pooled | 500 | GS | 0.49293 | 0.10526 | 0.880597 |
| NEUROCON | PPMI | svm_linear | 500 | GS | 0.500393 | 0.10526 | 0.895522 |
| NEUROCON | PPMI | gnb_pooled | 5000 | GS | 0.522977 | 0.1579 | 0.88806 |
| NEUROCON | PPMI | svm_linear | 5000 | GS | 0.538885 | 0.42105 | 0.656716 |
| NEUROCON | TaoWu | gnb_pooled | 10 | GS | 0.5 | 0.2 | 0.8 |
| NEUROCON | TaoWu | svm_linear | 10 | GS | 0.425 | 0.15 | 0.7 |
| NEUROCON | TaoWu | gnb_pooled | 50 | GS | 0.5 | 0.4 | 0.6 |
| NEUROCON | TaoWu | svm_linear | 50 | GS | 0.4 | 0.05 | 0.75 |
| NEUROCON | TaoWu | gnb_pooled | 100 | GS | 0.475 | 0.45 | 0.5 |
| NEUROCON | TaoWu | svm_linear | 100 | GS | 0.425 | 0 | 0.85 |
| NEUROCON | TaoWu | gnb_pooled | 500 | GS | 0.45 | 0.6 | 0.3 |
| NEUROCON | TaoWu | svm_linear | 500 | GS | 0.6 | 0.25 | 0.95 |
| NEUROCON | TaoWu | gnb_pooled | 5000 | GS | 0.375 | 0.55 | 0.2 |
| NEUROCON | TaoWu | svm_linear | 5000 | GS | 0.575 | 0.15 | 1 |
| PPMI | NEUROCON | gnb_pooled | 10 | GS | 0.623656 | 0.58065 | 0.666667 |
| PPMI | NEUROCON | svm_linear | 10 | GS | 0.499403 | 0.25807 | 0.740741 |
| PPMI | NEUROCON | gnb_pooled | 50 | GS | 0.549881 | 0.45161 | 0.648148 |
| PPMI | NEUROCON | svm_linear | 50 | GS | 0.488053 | 0.16129 | 0.814815 |
| PPMI | NEUROCON | gnb_pooled | 100 | GS | 0.462366 | 0.25807 | 0.666667 |
| PPMI | NEUROCON | svm_linear | 100 | GS | 0.51135 | 0.09677 | 0.925926 |
| PPMI | NEUROCON | gnb_pooled | 500 | GS | 0.595878 | 0.58065 | 0.611111 |
| PPMI | NEUROCON | svm_linear | 500 | GS | 0.502091 | 0.09677 | 0.907407 |
| PPMI | NEUROCON | gnb_pooled | 5000 | GS | 0.616487 | 0.67742 | 0.555556 |
| PPMI | NEUROCON | svm_linear | 5000 | GS | 0.423536 | 0.03226 | 0.814815 |
| PPMI | TaoWu | gnb_pooled | 10 | GS | 0.525 | 0.45 | 0.6 |
| PPMI | TaoWu | svm_linear | 10 | GS | 0.575 | 0.4 | 0.75 |
| PPMI | TaoWu | gnb_pooled | 50 | GS | 0.45 | 0.45 | 0.45 |
| PPMI | TaoWu | svm_linear | 50 | GS | 0.575 | 0.3 | 0.85 |
| PPMI | TaoWu | gnb_pooled | 100 | GS | 0.575 | 0.45 | 0.7 |
| PPMI | TaoWu | svm_linear | 100 | GS | 0.65 | 0.35 | 0.95 |
| PPMI | TaoWu | gnb_pooled | 500 | GS | 0.475 | 0.35 | 0.6 |
| PPMI | TaoWu | svm_linear | 500 | GS | 0.6 | 0.2 | 1 |
| PPMI | TaoWu | gnb_pooled | 5000 | GS | 0.55 | 0.55 | 0.55 |
| PPMI | TaoWu | svm_linear | 5000 | GS | 0.575 | 0.2 | 0.95 |
| TaoWu | NEUROCON | gnb_pooled | 10 | GS | 0.459976 | 0.29032 | 0.62963 |
| TaoWu | NEUROCON | svm_linear | 10 | GS | 0.434588 | 0.25807 | 0.611111 |
| TaoWu | NEUROCON | gnb_pooled | 50 | GS | 0.413978 | 0.16129 | 0.666667 |
| TaoWu | NEUROCON | svm_linear | 50 | GS | 0.446237 | 0.22581 | 0.666667 |
| TaoWu | NEUROCON | gnb_pooled | 100 | GS | 0.434588 | 0.25807 | 0.611111 |
| TaoWu | NEUROCON | svm_linear | 100 | GS | 0.441458 | 0.29032 | 0.592593 |
| TaoWu | NEUROCON | gnb_pooled | 500 | GS | 0.436679 | 0.35484 | 0.518519 |
| TaoWu | NEUROCON | svm_linear | 500 | GS | 0.443548 | 0.3871 | 0.5 |
| TaoWu | NEUROCON | gnb_pooled | 5000 | GS | 0.408901 | 0.35484 | 0.462963 |
| TaoWu | NEUROCON | svm_linear | 5000 | GS | 0.452808 | 0.3871 | 0.518519 |
| TaoWu | PPMI | gnb_pooled | 10 | GS | 0.536332 | 0.73684 | 0.335821 |
| TaoWu | PPMI | svm_linear | 10 | GS | 0.513354 | 0.57895 | 0.447761 |
| TaoWu | PPMI | gnb_pooled | 50 | GS | 0.520228 | 0.42105 | 0.619403 |
| TaoWu | PPMI | svm_linear | 50 | GS | 0.388845 | 0.21053 | 0.567164 |
| TaoWu | PPMI | gnb_pooled | 100 | GS | 0.512962 | 0.47368 | 0.552239 |
| TaoWu | PPMI | svm_linear | 100 | GS | 0.535546 | 0.52632 | 0.544776 |
| TaoWu | PPMI | gnb_pooled | 500 | GS | 0.630204 | 0.89474 | 0.365672 |
| TaoWu | PPMI | svm_linear | 500 | GS | 0.532011 | 0.57895 | 0.485075 |
| TaoWu | PPMI | gnb_pooled | 5000 | GS | 0.577965 | 0.89474 | 0.261194 |
| TaoWu | PPMI | svm_linear | 5000 | GS | 0.562451 | 0.73684 | 0.38806 |
| NEUROCON | PPMI | gnb_pooled | 10 | bandpass | 0.519049 | 0.10526 | 0.932836 |
| NEUROCON | PPMI | svm_linear | 10 | bandpass | 0.437156 | 0.1579 | 0.716418 |
| NEUROCON | PPMI | gnb_pooled | 50 | bandpass | 0.459348 | 0.10526 | 0.813433 |
| NEUROCON | PPMI | svm_linear | 50 | bandpass | 0.462883 | 0.05263 | 0.873134 |
| NEUROCON | PPMI | gnb_pooled | 100 | bandpass | 0.467203 | 0.21053 | 0.723881 |
| NEUROCON | PPMI | svm_linear | 100 | bandpass | 0.440888 | 0.1579 | 0.723881 |
| NEUROCON | PPMI | gnb_pooled | 500 | bandpass | 0.425766 | 0.10526 | 0.746269 |
| NEUROCON | PPMI | svm_linear | 500 | bandpass | 0.481932 | 0.1579 | 0.80597 |
| NEUROCON | PPMI | gnb_pooled | 5000 | bandpass | 0.414375 | 0.05263 | 0.776119 |
| NEUROCON | PPMI | svm_linear | 5000 | bandpass | 0.497054 | 0.21053 | 0.783582 |
| NEUROCON | TaoWu | gnb_pooled | 10 | bandpass | 0.475 | 0.15 | 0.8 |
| NEUROCON | TaoWu | svm_linear | 10 | bandpass | 0.525 | 0.35 | 0.7 |
| NEUROCON | TaoWu | gnb_pooled | 50 | bandpass | 0.475 | 0.35 | 0.6 |
| NEUROCON | TaoWu | svm_linear | 50 | bandpass | 0.5 | 0.25 | 0.75 |
| NEUROCON | TaoWu | gnb_pooled | 100 | bandpass | 0.525 | 0.45 | 0.6 |
| NEUROCON | TaoWu | svm_linear | 100 | bandpass | 0.575 | 0.35 | 0.8 |
| NEUROCON | TaoWu | gnb_pooled | 500 | bandpass | 0.475 | 0.35 | 0.6 |
| NEUROCON | TaoWu | svm_linear | 500 | bandpass | 0.7 | 0.4 | 1 |
| NEUROCON | TaoWu | gnb_pooled | 5000 | bandpass | 0.5 | 0.5 | 0.5 |
| NEUROCON | TaoWu | svm_linear | 5000 | bandpass | 0.65 | 0.35 | 0.95 |
| PPMI | NEUROCON | gnb_pooled | 10 | bandpass | 0.448626 | 0.19355 | 0.703704 |
| PPMI | NEUROCON | svm_linear | 10 | bandpass | 0.464456 | 0.35484 | 0.574074 |
| PPMI | NEUROCON | gnb_pooled | 50 | bandpass | 0.469235 | 0.29032 | 0.648148 |
| PPMI | NEUROCON | svm_linear | 50 | bandpass | 0.432796 | 0.03226 | 0.833333 |
| PPMI | NEUROCON | gnb_pooled | 100 | bandpass | 0.459976 | 0.29032 | 0.62963 |
| PPMI | NEUROCON | svm_linear | 100 | bandpass | 0.483572 | 0.09677 | 0.87037 |
| PPMI | NEUROCON | gnb_pooled | 500 | bandpass | 0.485364 | 0.32258 | 0.648148 |
| PPMI | NEUROCON | svm_linear | 500 | bandpass | 0.488351 | 0.03226 | 0.944444 |
| PPMI | NEUROCON | gnb_pooled | 5000 | bandpass | 0.476105 | 0.32258 | 0.62963 |
| PPMI | NEUROCON | svm_linear | 5000 | bandpass | 0.481481 | 0 | 0.962963 |
| PPMI | TaoWu | gnb_pooled | 10 | bandpass | 0.475 | 0.2 | 0.75 |
| PPMI | TaoWu | svm_linear | 10 | bandpass | 0.55 | 0.5 | 0.6 |
| PPMI | TaoWu | gnb_pooled | 50 | bandpass | 0.5 | 0.4 | 0.6 |
| PPMI | TaoWu | svm_linear | 50 | bandpass | 0.475 | 0.15 | 0.8 |
| PPMI | TaoWu | gnb_pooled | 100 | bandpass | 0.475 | 0.45 | 0.5 |
| PPMI | TaoWu | svm_linear | 100 | bandpass | 0.45 | 0.2 | 0.7 |
| PPMI | TaoWu | gnb_pooled | 500 | bandpass | 0.475 | 0.4 | 0.55 |
| PPMI | TaoWu | svm_linear | 500 | bandpass | 0.525 | 0.1 | 0.95 |
| PPMI | TaoWu | gnb_pooled | 5000 | bandpass | 0.45 | 0.45 | 0.45 |
| PPMI | TaoWu | svm_linear | 5000 | bandpass | 0.55 | 0.1 | 1 |
| TaoWu | NEUROCON | gnb_pooled | 10 | bandpass | 0.543309 | 0.29032 | 0.796296 |
| TaoWu | NEUROCON | svm_linear | 10 | bandpass | 0.515532 | 0.29032 | 0.740741 |
| TaoWu | NEUROCON | gnb_pooled | 50 | bandpass | 0.515532 | 0.29032 | 0.740741 |
| TaoWu | NEUROCON | svm_linear | 50 | bandpass | 0.364994 | 0.32258 | 0.407407 |
| TaoWu | NEUROCON | gnb_pooled | 100 | bandpass | 0.499403 | 0.25807 | 0.740741 |
| TaoWu | NEUROCON | svm_linear | 100 | bandpass | 0.41129 | 0.32258 | 0.5 |
| TaoWu | NEUROCON | gnb_pooled | 500 | bandpass | 0.506272 | 0.29032 | 0.722222 |
| TaoWu | NEUROCON | svm_linear | 500 | bandpass | 0.506272 | 0.29032 | 0.722222 |
| TaoWu | NEUROCON | gnb_pooled | 5000 | bandpass | 0.503883 | 0.32258 | 0.685185 |
| TaoWu | NEUROCON | svm_linear | 5000 | bandpass | 0.490143 | 0.25807 | 0.722222 |
| TaoWu | PPMI | gnb_pooled | 10 | bandpass | 0.568932 | 0.47368 | 0.664179 |
| TaoWu | PPMI | svm_linear | 10 | bandpass | 0.531422 | 0.42105 | 0.641791 |
| TaoWu | PPMI | gnb_pooled | 50 | bandpass | 0.607031 | 0.68421 | 0.529851 |
| TaoWu | PPMI | svm_linear | 50 | bandpass | 0.480754 | 0.84211 | 0.119403 |
| TaoWu | PPMI | gnb_pooled | 100 | bandpass | 0.621956 | 0.68421 | 0.559701 |
| TaoWu | PPMI | svm_linear | 100 | bandpass | 0.454635 | 0.84211 | 0.067164 |
| TaoWu | PPMI | gnb_pooled | 500 | bandpass | 0.652003 | 0.73684 | 0.567164 |
| TaoWu | PPMI | svm_linear | 500 | bandpass | 0.473095 | 0.78947 | 0.156716 |
| TaoWu | PPMI | gnb_pooled | 5000 | bandpass | 0.633346 | 0.73684 | 0.529851 |
| TaoWu | PPMI | svm_linear | 5000 | bandpass | 0.547722 | 0.78947 | 0.30597 |

## *Consensus NMF clustering*

For a more direct graphical depiction of the heterogeneity of the functional connectomes of the PD patient scans, we have applied *consensus NMF clustering* [Brunet et al., 2004] for a progressively increasing number of clusters *k*=2,…,18 (Supplementary Fig 3). The Figure depicts the symmetric consensus co-clustering matrices for the PD scans from the NEUROCON dataset. Note that besides the consistent grouping of the replicate scan pairs for each patient, it is difficult to single out an optimal number of clusters *k*.

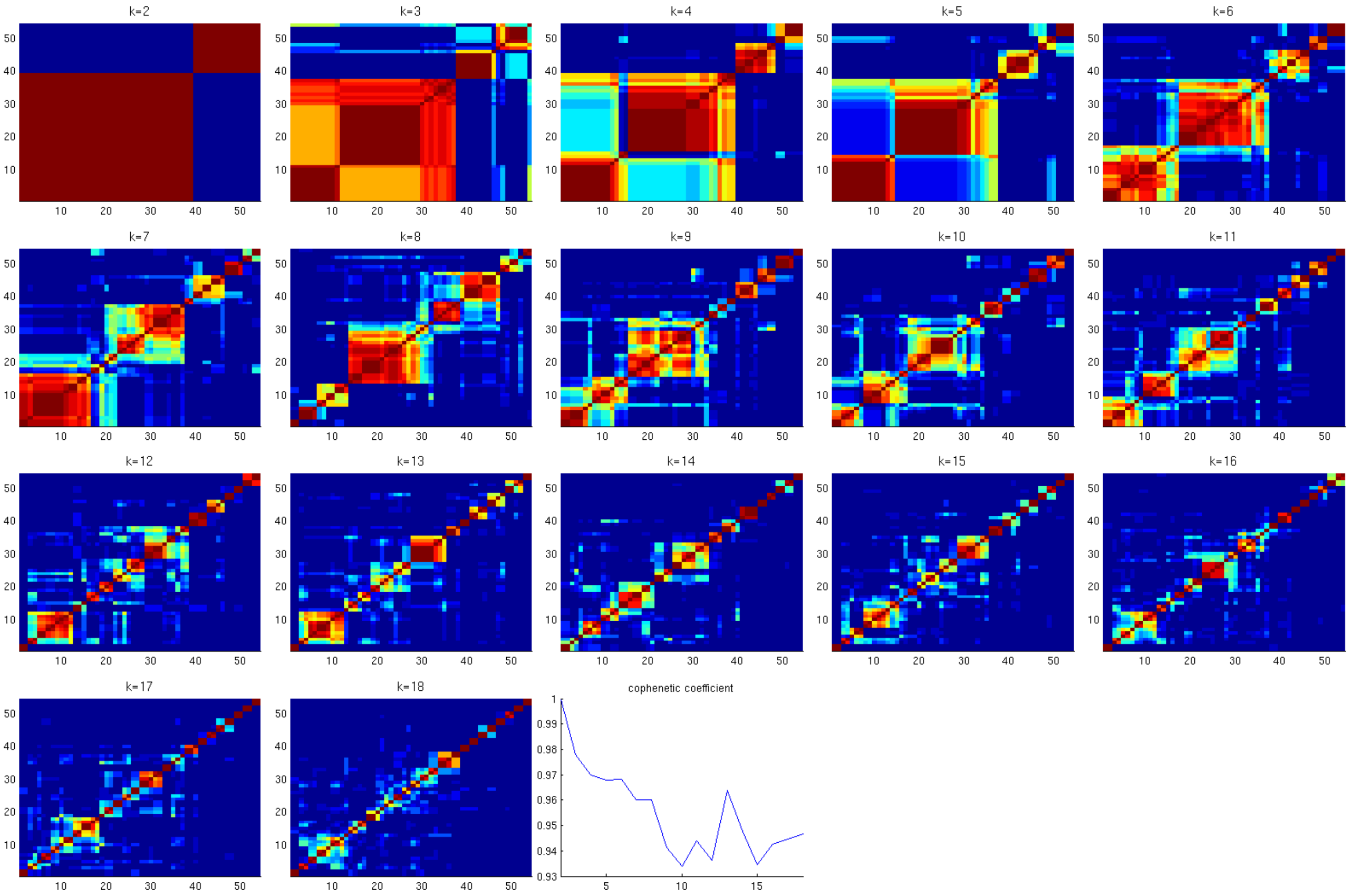

**Supplementary Fig 3. Consensus NMF clustering of functional connectomes of PD patient scans from the NEUROCON dataset.**